\documentclass[aps,10pt]{revtex4-1}

\usepackage{graphicx} 
\usepackage{epsfig}
\usepackage{amsmath}
\usepackage{amssymb}
\usepackage{xcolor}

\def\ov#1{\overline{#1}}

\def\vb#1{\mbox{\boldmath$#1$}}
\def\pd#1#2{\frac{\partial #1}{\partial #2}}

\def\wh#1{\widehat{#1}}
\def\bdot{\,\vb{\cdot}\,}
\def\btimes{\,\vb{\times}\,}

\def\cal#1{\mathcal{#1}}

\begin{document}

\title{Nonlinear Schr\"{o}dinger Equation on a closed 3D Elastica Knot}

\author{Alain J. Brizard}

\affiliation{Department of Physics, Saint Michael's College, Colchester, VT 05439, USA}

\date{\today}

\begin{abstract}
The present work introduces a new relation between closed elastica knots with the solution of the nonlinear Schr\"{o}dinger equation (NLSE) through the Hasimoto transformation. An elastica knot is defined in terms of the Frenet-Serret curvature $\kappa(s,t)$ as a function of the arclength $s$ along the spatial curve  ${\bf r}(s,t)$ at a fixed time $t$, which is a solution of the curvature differential equation $\partial^{2}_{s}\kappa(s,t) = -\;\kappa^{3}/2 + k_{0}^{4}\tau_{0}^{2}\;\kappa^{-3} + \lambda\,k_{0}^{2}\kappa/2$ that is obtained from a variational principle that minimizes the bending energy of the spatial curve under the constraint of a constant curve length. Here, the Frenet-Serret torsion $\tau(s,t)$ satisfies the conservation law $\kappa^{2}(s,t)\,\tau(s,t) \equiv k_{0}^{2}\,\tau_{0}$, while the constant of integration $\lambda$ is associated with the constrained variation. After briefly reviewing the Hasimoto transformation from a space curve ${\bf r}(s,t)$ to the nonlinear Schr\"{o}dinger equation (NLSE) $-\,iD^{-1}\partial_{t}\psi = \partial^{2}_{s}\psi + \frac{1}{2}\,|\psi|^{2}\psi$, where the constant $D$ has units of fluid circulation (m$^{2}$/sec), we show how the traveling-wave solution $\psi(s,t) = \Psi(s_{t} \equiv s - c\,t) \equiv \kappa(s_{t})\;\exp[i\theta(s_{t})]$ is mapped onto the curvature equation for an elastica knot, with $\theta^{\prime}(s_{t}) \equiv c/(2D) + k_{0}^{2}\tau_{0}/\kappa^{2}(s_{t})$ and the elastica-knot constant $k_{0}^{2}\lambda = -\frac{1}{2}\,(c/D)^{2}$ expressed in terms of the traveling-wave NLSE parameters $(c,D)$. The constraint of a closed 3D elastica knot imposes spatial periodicity conditions that introduce a unique set of knot parameters for which the NLSE traveling wave can exist. The present work shows that the NLSE traveling-wave solution on a closed elastica knot requires an extension of the classical elastica-knot parameter space.
\end{abstract}

\maketitle

\section{Introduction} 

Observations of planar and non-planar vortex filaments, as well as elastic rods tied in knots, have provided ample fascination for several centuries. Their mathematical analysis was initiated by Euler, Bernoulli, Lagrange, and Kirchhoff, among others \cite{Nizette_1999,Singer_2008,Matsutani_2010}.

There is also a rich history of the connection between the motion of non-stretching vortex filaments (i.e., closed spatial curves) \cite{Betchov_1965,Ricca_1991,Nakayama_1992,Ricca_1996,Kleckner_2013}, which are characterized by their Frenet-Serret curvature and torsion \cite{Nizette_1999}, and the nonlinear Schr\"{o}dinger equation (NLSE) through the Hasimoto transformation  \cite{Hasimoto_1971,Hasimoto_1972,Lamb_1976,Lamb_1977,Kida_1981,Hasimoto_1988,Balakrishnan_1999}. In particular, Zakharov and Shabat \cite{ZS_1972} demonstrated that the NLSE possesses solitary wave solutions, in analogy with other integrable infinite-dimensional Hamiltonian systems \cite{Ablowitz_2008}. The NLSE has appeared in many important areas of physics \cite{Zakharov_1968,Dewar_1972,KBD_1977,KBD_1978,Peregrine_1983,Uby_1995, Sulem_2007,KBD_2008,Salman_2013, AM_2017,Karjanto_2019,Carter_2020}, and its connection to the Frenet-Serret curvature and torsion of arbitrary spatial curves continues to attract interest \cite{Smondyrev_1995,Salman_2014}.

A special category of spatial curves that have generated significant interest in many areas of mathematics and physics are the 3D elastica curves \cite{Langer_Singer_1984,LS_1984,Langer_Singer_1996,Barros_2018}, which generalize the 2D (planar) elastica curves \cite{Matsutani_2010}. These 3D elastica curves can either be helical curves or torus knots \cite{Langer_Singer_1984}. For 3D helical curves, the solution $\psi(s,t)$ of the NLSE must take into account behavior at infinity, i.e., unless $\psi(s,t)$ is periodic in $s$, it must vanish as $s \rightarrow \pm\infty$. Elastica curves on torus knots, on the other hand, which are built from spatial periodicity conditions, may either be open or closed.

\subsection{Main results and motivation for the present work}

The main results of this work are as follows. First, we extend the parameter space for a closed elastica-knot spatial curve. Second, we show that an NLSE traveling-wave solution on a closed elastica knot, which is connected to the Frenet-Serret curvature and torsion of a spatial curve by the Hasimoto transformation \cite{Hasimoto_1972}, exists only in this extended elastica-knot parameter space. In fact, Langer and Singer \cite{LS_1984} (see Table 2.8) explicitly forbids an NLSE traveling-wave solution along an elastica knot in the parameter range for which $2 + \gamma^{2} - \nu^{2} < 0$, where $\gamma \equiv c/(k_{0}D)$ and $\nu \equiv 2\tau_{0}/k_{0}$ correspond to a dimensionless wave group velocity and dimensionless torsion, respectively. The present work shows that NLSE traveling-wave solutions indeed exist for elastica knots, but only if the elastica-knot parameter range is extended to include the condition $2 + \gamma^{2} - \nu^{2} < 0$ [see Eq.~(\ref{eq:m_forbidden}) and Fig.~\ref{fig:Q0_m}]. Third, two equivalent representations of the spatial curves corresponding to closed elastica knots are given in terms of the Jacobi and Weierstrass elliptic functions. We note that the Weierstrass elliptic representation allows us to express new results for the NLSE traveling-wave solutions on closed elastica knots in closed form. An explicit example for a spatial curve associated with a closed elastica knot in the extended parameter range, which an NLSE traveling-wave solution exists, is given.

The primary result, and a major motivation for the present work, involves the extension of the elastica parameter range, through a change in notation involving the elliptic functions and integrals. On the one hand, the mathematical notation \cite{Lawden,NIST_Chap22} used by Langer and Singer \cite{Langer_Singer_1984,LS_1984}, for example, considers the following notation for the generic Jacobi function ${\rm pq}(x,p)$, where the argument $x$ is real and the modulus $p$ is assumed to be in the {\it classical} range $0 \leq p \leq 1$. 

The present work, on the other hand, uses the conventional notation \cite{AS} for the generic Jacobi elliptic functions ${\rm pq}(z|m)$, where the argument $z$ may be complex valued while the parameter $- \infty < m \leq 1$ may be negative. When $0 \leq m \leq 1$, we find the identity between the two notations: ${\rm pq}(x,p) \equiv {\rm pq}(x| m=p^{2})$. For the extended range $m < 0$, however, the classical modulus $p \equiv i\,\sqrt{|m|}$ becomes imaginary, which is forbidden from the standard elastica-knot analysis \cite{Langer_Singer_1984,LS_1984,Langer_Singer_1996,Barros_2018}. In the present work, we use the standard transformations \cite{AS} (summarized in App.~\ref{sec:Jacobi}) from the extended range $m < 0$ to the classical range $0 < n \equiv 1 - n' < 1$:
\begin{equation} 
    {\rm pq}\left(\left.\begin{array}{c}
    x \\
    i\,y
    \end{array}\right|m\right) \;\equiv\; \left\{ \begin{array}{l}
    \ov{\rm pq}(x/\sqrt{n'}|n) \\
    \\
    \ov{\rm rs}(y/\sqrt{n'}|n')
    \end{array} \right.
    \label{eq:transform_J}
\end{equation}
where $\ov{\rm pq}$ and $\ov{\rm rs}$ are Jacobi functions of $x$ or $y$, and $(n,n')$ are functions of $m$ such that the new Jacobi parameter $n$ falls in the classical range $0 < n < 1$. Following the work of Pfefferl\'{e} {\it et al.} \cite{Pfefferle_2018} on the non-planar elastica representations of the magnetic axis of a stellarator, Brizard and Pfefferl\'{e} \cite{AJB_DP} showed that, for each elastica knot in the classical range $0 \leq m \leq 1$, there exists an equivalent elastica knot in the extended range $m \leq 0$, i.e., $0 \leq n(m) \leq 1$, with identical global elastica-knot properties (see App.~\ref{sec:equivalent} for a summary). For example, the normalized total curvature of a classical elastica knot with $n = 1/2$ is equal to the normalized total curvature of a extended elastica knot with $m = -1$, even though their respective spatial curves are different.

Finally, we note that all figures presented in this work are produced with {\sf Mathematica}, which follows the conventional notation \cite{AS} for the Jacobi elliptic functions and elliptic integrals. Hence, expressions can be calculated and plotted continuously throughout the parameter range $-\infty < m \leq 1$.

\subsection{Organization}

The remainder of the paper is organized as follows. In Sec.~\ref{sec:sec_2}, we derive the curvature equation (\ref{eq:kappa_pp}) for an elastica knot from a constrained variational principle (\ref{eq:delta_F}) that minimizes the total squared curvature under the constraint of constant total knot length. This curvature equation for $\kappa(s)$ takes into account the torsion-curvature conservation law $\kappa(s)\tau(s) = k_{0}^{2}\tau_{0}$, where $k_{0} \equiv \kappa(0)$ and $\tau_{0} \equiv \tau(0)$, and introduces an arbitrary integration constant $\lambda$ associated with the Lagrange multiplier used in the constrained variational principle. In Sec.~\ref{sec:sec_3}, we briefly review the Hasimoto transformation \cite{Hasimoto_1972} from the Frenet-Serret curvature $\kappa(s,t)$ and torsion $\tau(s,t)$ of a spatial curve ${\bf r}(s,t)$ to the solution $\psi(s,t) \equiv \kappa(s,t)\;\exp[i\int_{0}^{s}\tau(s',t)\,ds']$ of the nonlinear Schr\"{o}dinger equation (\ref{eq:NLSE}). In Sec.~\ref{sec:sec_4}, we show that the traveling-wave solution $\kappa(s_{t})\,\exp[i\,\theta(s_{t})]$ of the NLSE (\ref{eq:NLSE}), where $s_{t} \equiv s - c\,t$ is defined in terms of the constant wave speed $c$, matches the elastica-knot curvature equation (\ref{eq:kappa_pp}) if $\theta^{\prime}(s_{t}) = c/(2D) + k_{0}^{2}\tau_{0}/\kappa^{2}(s_{t})$ and the elastica-knot constant $k_{0}^{2}\lambda = -\,c^{2}/(2D^{2})$ is expressed in terms of the traveling-wave NLSE parameters $(c,D)$. 

In Sec.~\ref{sec:sec_5}, we solve the elastica-knot curvature equation (\ref{eq:kappa_elastica}) in terms of the Jacobi elliptic functions (\ref{eq:elastica_J}) and the Weierstrass elliptic functions (\ref{eq:W_kappa}), which are expressed in terms of two elastica-knot parameters $(m,q_{0})$, and throughout this work, the conventional notation \cite{AS} is used for elliptic functions and integrals. The requirement of real torsion constant $\tau_{0}$ and a real wave speed $c$ identifies two regions in parameter space $(m,q_{0})$, which are shown in Fig.~\ref{fig:gammanu_2} as regions bounded by three dashed lines a, b, and c. While a part of region I is located within the classical parameter range $0 < m < 1$, region II is entirely located within the extended parameter range $m < -1$. Next, in Sec.~\ref{sec:sec_6}, we construct a closed 3D spatial curve from our elastica-knot solution (\ref{eq:elastica_J}), which introduces constraints in elastica-knot parameter space $(m,q_{0})$, where $m = m_{c}$ and $q_{0} = Q(m_{c})$, which is defined in Eq.~(\ref{eq:q_LS}) in terms of complete elliptic integrals. In Sec.~\ref{sec:sec_7}, we derive a periodic solution to the NLSE (\ref{eq:NLSE}) through a direct comparison with the Lam\'{e} equation \cite{Ince_1940,Erdelyi_1941,NIST_Chap29}. We also give the Jacobi and Weierstrass elliptic expressions for the traveling-wave solution $\psi(s,t) \equiv \kappa(s_{t})\;\exp[i\theta(s_{t})]$ expressed in terms of the elastica-knot parameters $(m,q_{0})$. We conclude our work in Sec.~\ref{sec:sec_8}, while we present additional details in support of our presentation in Apps.~\ref{sec:Jacobi}-\ref{sec:planar}.

\section{\label{sec:sec_2}Elastica Knot Curvature Equation}

Elastica knots \cite{Langer_Singer_1984,LS_1984} are spatial curves that minimize the bending energy represented by the constrained curvature functional \cite{AJB_DP}
\begin{equation}
{\cal F}_{\Lambda}[{\bf r}] = \frac{1}{2}\;\int_{a}^{b} \left[ |{\bf r}^{\prime\prime}(s)|^{2} \;+\frac{}{} \Lambda(s)\;\left(|{\bf r}^{\prime}(s)|^{2} - 1 \right) \right] ds,
\label{eq:F_Lambda}
\end{equation}
where the curve ${\bf r}(s)$ is parameterized by the arclength position $s$ along the curve, and the function $\Lambda(s)$ serves as a Lagrange multiplier associated with the constraint $|{\bf r}^{\prime}(s)|^{2} = 1$ (since $ds^{2} \equiv |d{\bf r}|^{2}$).

The Frenet-Serret triad $(\wh{\sf t},\wh{\sf n},\wh{\sf b})$ is introduced through the Frenet-Serret equations \cite{Brizard_2015}
\begin{eqnarray}
    \partial\wh{\sf t}/\partial s &=& \kappa(s)\;\wh{\sf n}, \label{eq:t_FS} \\
    \partial\wh{\sf n}/\partial s &=& \tau(s)\;\wh{\sf b} \;-\; \kappa(s)\;\wh{\sf t}, \label{eq:n_FS} \\
    \partial\wh{\sf b}/\partial s &=& -\;\tau(s)\;\wh{\sf n}, \label{eq:b_FS}
\end{eqnarray}
which are expressed in terms of the curvature $\kappa$ and the torsion $\tau$, where $\wh{\sf t}$ denotes the tangent unit vector to the curve, $\wh{\sf n}$ denotes the normal unit vector (directed toward the local center of curvature), and $\wh{\sf b} \equiv \wh{\sf t}\btimes\wh{\sf n}$ denotes the binormal unit vector. From these Frenet-Serret equations, we obtain the partial derivatives \cite{Kida_1981}
\begin{equation}
\left. \begin{array}{rcl}
{\bf r}^{\prime} & = & \wh{\sf t} \\
{\bf r}^{\prime\prime} & = & \kappa\;\wh{\sf n} \\
{\bf r}^{\prime\prime\prime} & = & \kappa^{\prime}\;\wh{\sf n} \;+\; \kappa\,\left( \tau\;\wh{\sf b} \;-\; \kappa\;\wh{\sf t}\right)
\end{array} \right\},
\label{eq:r_primes}
\end{equation}
from which we obtain the definitions for the Frenet-Serret curvature $\kappa(s) \equiv |{\bf r}^{\prime\prime}|$ and the Frenet-Serret torsion $\tau(s) \equiv 
({\bf r}^{\prime}\btimes{\bf r}^{\prime\prime}\bdot{\bf r}^{\prime\prime\prime})/\kappa^{2}(s)$ (i.e., a non-planar curve has finite torsion). We note that, while the curvature is always positive, the torsion may be positive, negative, or zero.

The Euler equation for the curve ${\bf r}(s)$ is obtained from the first variation of the curvature functional (\ref{eq:F_Lambda}) \cite{Brizard_2015}:
\begin{eqnarray}
\delta{\cal F}_{\Lambda} &\equiv& \left(\frac{d}{d\epsilon}{\cal F}_{\Lambda}[{\bf r} + \epsilon\,\delta{\bf r}]\right)_{\epsilon = 0} \;=\; \int_{a}^{b}\left( {\bf r}^{\prime\prime}\bdot\delta{\bf r}^{\prime\prime} +\frac{}{} \Lambda\;{\bf r}^{\prime}\bdot\delta{\bf r}^{\prime}\right) ds \;=\; \int_{a}^{b} \delta{\bf r}\bdot\frac{d{\bf W}}{ds}ds \;\equiv\; 0,
\label{eq:delta_F}
\end{eqnarray}
where the variation $\delta{\bf r}$ and its first derivative $\delta{\bf r}^{\prime}$ are assumed to vanish at the end points $s = a$ and $s = b$, and the vector
\begin{eqnarray}
{\bf W}(s) & \equiv & {\bf r}^{\prime\prime\prime} - \Lambda\;{\bf r}^{\prime} = \kappa^{\prime}\;\wh{\sf n} + \kappa\,\tau\;\wh{\sf b} - \left(\kappa^{2} + \Lambda\right)\;\wh{\sf t}
 \label{eq:W_def}
 \end{eqnarray}
is written in terms of Eq.~(\ref{eq:r_primes}) and the Lagrange multiplier $\Lambda(s)$. 

When the first variation (\ref{eq:delta_F}) vanishes for arbitrary variations $\delta{\bf r}$ (subject to vanishing boundary conditions), we obtain the Euler equation relating the curvature $\kappa$ and the torsion $\tau$ for the curve ${\bf r}(s)$:
 \begin{eqnarray}
 0 = \frac{d{\bf W}}{ds} & = & -\;\left( 3\,\kappa\,\kappa^{\prime} \;+\frac{}{} \Lambda^{\prime}\right)\wh{\sf t} + \left( 2\,\kappa^{\prime}\;\tau \;+\frac{}{} \kappa\;\tau^{\prime}\right) \wh{\sf b} \;+\; \left[\kappa^{\prime\prime} - \kappa \left( \kappa^{2} \;+\frac{}{}\tau^{2} + \Lambda\right) \right] \wh{\sf n},
 \label{eq:Euler-W}
 \end{eqnarray}
 where each component vanishes identically. The $\wh{\sf t}$-component of Eq.~(\ref{eq:Euler-W}) yields $(\frac{3}{2}\,\kappa^{2} + \Lambda)^{\prime} = 0$, from which we obtain a solution for the Lagrange multiplier
 \begin{equation}
 \Lambda(s) \;\equiv\; -\;\frac{3}{2}\;\kappa^{2}(s) \;+\; \frac{1}{2}\,\lambda\,k_{0}^{2},
 \label{eq:Lambda_eq}
 \end{equation}
 where $\lambda$ denotes a dimensionless constant of integration (initially assumed to be $-\infty < \lambda < \infty$) and the curvature parameter $k_{0}$ is defined as $k_{0} \equiv \kappa(0)$. The $\wh{\sf b}$-component of Eq.~(\ref{eq:Euler-W}) yields the conservation law $(\kappa^{2}\,\tau)^{\prime} = 0$, from which we obtain the torsion constraint
 \begin{equation}
 \kappa^{2}(s)\;\tau(s) \;=\; {\bf r}^{\prime}\btimes{\bf r}^{\prime\prime}\bdot{\bf r}^{\prime\prime\prime} \;\equiv\; k_{0}^{2}\,\tau_{0},
 \label{eq:tau_eq}
 \end{equation}
 where the torsion parameter $\tau_{0} $ is defined as $\tau_{0} \equiv \tau(0)$. Substituting Eqs.~(\ref{eq:Lambda_eq})-(\ref{eq:tau_eq}) into Eq.~(\ref{eq:W_def}), the components of ${\bf W}$ become functions of $\kappa$ and  $\kappa^{\prime}$:
 \begin{equation}
{\bf W} \;=\; \kappa^{\prime}\;\wh{\sf n} \;+\; k_{0}^{2}\tau_{0}\,\kappa^{-1}\;\wh{\sf b} \;+\; \frac{1}{2}\,\left(\kappa^{2} \;-\;\lambda\,k_{0}^{2} \right)\wh{\sf t}.
 \label{eq:W_const}
 \end{equation}
 Lastly, the $\wh{\sf n}$-component of Eq.~(\ref{eq:Euler-W}) yields the curvature second-order ordinary differential equation
\begin{eqnarray}
    \kappa^{\prime\prime} & = & -\;\frac{1}{2}\,\kappa^{3} \;+\; 
    k_{0}^{4}\tau_{0}^{2}\;\kappa^{-3} \;+\; \frac{1}{2}\;\lambda\,
    k_{0}^{2}\kappa,
\label{eq:kappa_pp}
\end{eqnarray}
where we inserted the relations (\ref{eq:Lambda_eq})-(\ref{eq:tau_eq}). The elastica curvature equation (\ref{eq:kappa_pp}) is identical to the curvature equation presented by Langer and Singer \cite{Langer_Singer_1984}, where a different notation is used (i.e., $\lambda - 2G = \lambda\,k_{0}^{2}$ and $c = k_{0}^{2}\tau_{0}$ in our work). Here, the curvature solution $\kappa(s;\tau_{0},\lambda)$ is parametrized by the torsion constant $\tau_{0}$ and the Lagrange multiplier constant $\lambda$, while $k_{0}$ is an arbitrary curvature scale parameter.

\section{\label{sec:sec_3}Hasimoto Transformation}

In this Section, we briefly review the Hasimoto transformation \cite{Hasimoto_1972} that connects the curvature and torsion of a time-dependent spatial curve with the solution of the nonlinear Schr\"{o}dinger equation. Because of the extraordinary novelty of this connection, we present a complete derivation of this remarkable transformation.

First, we consider a space curve ${\bf r}(s,t)$, which is a function of spatial position $s$ along the curve at time $t$. Partial derivatives of ${\bf r}(s,t)$ with respect to $s$ and $t$ are expressed as  \cite{Hasimoto_1972}
\begin{eqnarray}
    \pd{\bf r}{s} &\equiv& \wh{\sf t}, \label{eq:X_s} \\
    \pd{\bf r}{t} &\equiv& D\;\pd{\bf r}{s} \btimes \frac{\partial^{2}{\bf r}}{\partial s^{2}} \;=\; D\;\wh{\sf t}\btimes\pd{\wh{\sf t}}{s} \;=\; D\,\kappa\;\wh{\sf b}, \label{eq:X_t} 
\end{eqnarray}
where the constant $D$ has units of fluid circulation (${\rm m}^{2}$/sec) when a vortex filament is considered. (Note: In contrast to Hasimoto's work \cite{Hasimoto_1972}, we retain physical units for a clearer physical perspective.) From a historical point of view, the curvature drift velocity (\ref{eq:X_t}) is known as the Betchov-Da Rios equation \cite{Betchov_1965,Ricca_1991,Ricca_1996} or {\it local induction} equation, which states that the filament velocity depends on the instantaneous local curvature $\kappa(s,t)$  and moves in the binormal direction $\wh{\sf b}$ (i.e., perpendicular to the local plane spanned by the tangent unit vector $\wh{\sf t}$ and the normal unit vector $\wh{\sf n}$). We point out an interesting analogy with the singular parallel guiding-center motion \cite{Cary_Brizard_2009} of a charged particle (with mass $M$ and charge $e$), where the magnetic moment invariant $\mu \equiv 0$ vanishes, so that the guiding-center velocity
\[ \frac{d{\bf r}}{dt} \;\equiv\; \frac{ds}{dt}\,\pd{\bf r}{s} \;+\; \pd{\bf r}{t} \;=\; v_{\|}\,\wh{\sf t} \;+\;
\frac{v_{\|}^{2}}{\Omega}\;\wh{\sf t}\btimes\pd{\wh{\sf t}}{s} \]
is represented as the sum of the guiding-center parallel velocity $v_{\|} = ds/dt$ along the magnetic field line ${\bf B} \equiv B\,\wh{\sf t}(s)$ and the magnetic curvature drift velocity. Here, we note that $v_{\|}$ is a constant of motion because of energy conservation, and the gyrofrequency $\Omega \equiv eB/M$ can be used to define the constant $D \equiv v_{\|}^{2}/\Omega$ in Eq.~(\ref{eq:X_t}).

Second, we note that Eqs.~(\ref{eq:n_FS})-(\ref{eq:b_FS}) can be written as
\[ \partial_{s}(\wh{\sf n} + i\,\wh{\sf b}) \;+\; i\tau\,(\wh{\sf n} + i\,\wh{\sf b}) \;=\; 
-\kappa\,\wh{\sf t}, \]
which can be rewritten as
\begin{equation}
    e^{-i\Theta} \pd{}{s}\left[ e^{i\Theta}\left(\wh{\sf n} + i\,\wh{\sf b}\right)\right] \;=\; -\,\kappa\;\wh{\sf t},
    \label{eq:LIE_N}
\end{equation}
where the integrating factor $\Theta$ is defined as an integral of the Frenet-Serret torsion along the arclength path from $0$ to $s$ at constant time $t$:
\begin{equation}
    \Theta(s,t) \;\equiv\; \int_{0}^{s}\tau(s',t)\;ds'.
\end{equation}
Hence, using the Hasimoto transformation  \cite{Hasimoto_1972}, we define the scalar and vector functions
\begin{eqnarray}
    \psi &\equiv& \kappa\;\exp(i\,\Theta), \label{eq:psi_def} \\
    {\bf N} &\equiv& \left(\wh{\sf n} + i\,\wh{\sf b}\right)\;\exp(i\,\Theta), \label{eq:N_def}
\end{eqnarray}
so that Eq.~(\ref{eq:LIE_N}) becomes
\begin{equation}
    \pd{\bf N}{s} \;=\; -\;\psi\;\wh{\sf t}. \label{eq:N_s}
\end{equation}
This equation formally connects the curvature and torsion of the space curve, represented by the vector function (\ref{eq:N_def}), with the complex-valued scalar function (\ref{eq:psi_def}), which will later be expressed as the solution of the nonlinear Schr\"{o}dinger equation.

\subsection{Nonlinear Schr\"{o}dinger equation}

We now proceed to evaluate the partial time derivatives of $\wh{\sf t}$ and ${\bf N}$. First, we immediately note that Eq.~(\ref{eq:t_FS}) can be expressed as 
\begin{equation} 
    \pd{\wh{\sf t}}{s} \;=\; \frac{\kappa}{2}\left[ \left(\wh{\sf n} + i\,\wh{\sf b}\right) \;+\; \left(\wh{\sf n} - i\,\wh{\sf b}\right)\right] \;\equiv\; {\rm Re}(\psi^{*}\,{\bf N}). \label{eq:t_s}
\end{equation}
Second, Eqs.~(\ref{eq:X_s})-(\ref{eq:X_t}) yield
\begin{eqnarray}
    \pd{\wh{\sf t}}{t} \;=\; \frac{\partial^{2}{\bf r}}{\partial s\partial t} &=& D\left( \kappa^{\prime}\,\wh{\sf b} \;-\; \kappa\,\tau\;\wh{\sf n}
    \right) \;\equiv\; -\;{\rm Im}\left(D\,\psi^{\prime}\;{\bf N}^{*}\right),
\end{eqnarray}
where a prime denotes a partial derivative with respect to $s$, with
\begin{equation}
    \psi^{\prime} \;\equiv\; \pd{\psi}{s} \;=\; \left(\frac{\kappa^{\prime}}{\kappa} \;+\; i\,\tau\right) \psi,
\end{equation}
which follows from the definition (\ref{eq:psi_def}). Third, by using the identities ${\bf N}\bdot{\bf N} = 0$, ${\bf N}\bdot{\bf N}^{*} = 2$, and ${\bf N}\bdot\wh{\sf t} = 0$, we obtain
\cite{Hasimoto_1972}
\begin{equation}
    \pd{\bf N}{t} \;=\; i\;\left(f\;{\bf N} \;-\; D\,\pd{\psi}{s}\;\wh{\sf t}\right), \label{eq:N_t}
\end{equation}
where $f$ is an unspecified function that will be determined later [see Eq.~(\ref{eq:nu_def})]. 

Finally, we use Eqs.~(\ref{eq:N_s}) and (\ref{eq:N_t}) to obtain
\begin{eqnarray*}
    \frac{\partial^{2}{\bf N}}{\partial s\partial t} &=& i \left[ \pd{f}{s}\,{\bf N} - D\kappa\,\pd{\psi}{s}\;\wh{\sf n} - \left(f\,\psi + D\frac{\partial^{2}\psi}{\partial s^{2}}\right)\wh{\sf t} \right], \\
    \frac{\partial^{2}{\bf N}}{\partial t\partial s} &=& -\;\pd{\psi}{t}\;\wh{\sf t} - \frac{iD}{2}\;\left(\psi\,\pd{\psi}{s}\;{\bf N}^{*} -
    \psi\,\pd{\psi^{*}}{s}\;{\bf N}\right).
\end{eqnarray*}
Since partial derivatives with respect to $s$ and $t$ commute, these two equations are identical, so that they have identical components when dotted into ${\bf N}$. Next, when dotted into ${\bf N}^{*}$, we obtain the identity
\[ 2\,\pd{f}{s} \;-\; D\,\psi^{*}\;\pd{\psi}{s} \;=\; D\,\psi\;
\pd{\psi^{*}}{s}, \]
which is solved as
\begin{equation}
    f \;\equiv\; \frac{D}{2}\;|\psi|^{2} \;=\; \frac{1}{2}\,D\,\kappa^{2},
    \label{eq:nu_def}
\end{equation}
where we used Eq.~(\ref{eq:psi_def}) and we ignored an arbitrary function of time. Lastly, when dotted into $\wh{\sf t}$, we obtain the (self-focusing) nonlinear Schr\"{o}dinger equation \cite{ZS_1972,Ablowitz_2008}
\begin{equation}
    -i\,D^{-1}\pd{\psi}{t} \;=\; \frac{\partial^{2}\psi}{\partial s^{2}} \;+\; \frac{1}{2}\,|\psi|^{2}\,\psi. 
    \label{eq:NLSE}
\end{equation}
We note that, from a quantum mechanical point of view, the constant $D = \hbar/(2M)$ in Eq.~(\ref{eq:NLSE}) can also be defined in terms of Planck's constant $\hbar$ and the mass $M$ of a hypothetical particle moving in the potential $V \equiv -\,\hbar^{2}\kappa^{2}/(4M)$. Other interpretations of $D$ depend on the primitive physical model on which the nonlinear Schr\"{o}dinger equation (\ref{eq:NLSE}) is derived (see Refs.~\cite{Zakharov_1968,Dewar_1972}, for example).

\subsection{\label{sec:simple_NLSE}Simple NLSE soliton solutions} 

We now explore a few simple solutions of the NLSE (\ref{eq:NLSE}), which are either separable solutions or traveling-wave envelope solitons. Beforehand, we note two invariance properties of the NLSE (\ref{eq:NLSE}). First, Eq.~(\ref{eq:NLSE}) is invariant under the scaling transformation $t \rightarrow a^{2}\,t$, $s \rightarrow a\,s$, and $\psi \rightarrow a^{-1}\psi$, where $a$ is an arbitrary constant, which implies that, if $\psi(s,t)$ is a solution of the NLSE (\ref{eq:NLSE}), then 
\begin{equation}
    \psi(s,t) \;\rightarrow\; a^{-1}\,\psi(a\,s, a^{2}\,t)
    \label{eq:NLSE_scale}
\end{equation} 
is also a solution.

Next, we consider the Galilean transformation $(s,t) \rightarrow (s_{t} \equiv s - v\,t,\; t)$, and the gauge transformation 
\begin{equation} 
\psi(s,t) \;\rightarrow\; \psi(s_{t},t)\,\exp[i\Theta(s_{t},t)], 
\label{eq:gauge}
\end{equation}
where $v$ is an arbitrary constant and the gauge phase $\Theta(s_{t},t)$ is a linear function of $(s_{t},t)$. We note that the NLSE (\ref{eq:NLSE}) is invariant under this transformation if the gauge phase is \cite{Zakharov_1968} $\Theta = (v/2D)\,(s_{t} + vt/2)$, which implies that if $\psi(s,t)$ is a solution of the NLSE (\ref{eq:NLSE}), then
\begin{equation}
    \psi(s,t) \;\rightarrow\; \psi(s_{t},t)\;\exp\left[ \frac{iv}{2D} \left(s_{t} \;+\; \frac{vt}{2}\right) \right] \;=\; \psi(s - vt, t) \;\exp\left[\frac{i\,v}{2D}\,\left( s \;-\;  \frac{v}{2}\,t\right)\right]
    \label{eq:gauge_NLSE}
\end{equation}
is also a solution.

\subsubsection{Simple separable NLSE solutions}

The simplest separable NLSE solutions include the time-dependent solution $\Psi(t) = k_{0}\,\exp(iD\,k_{0}^{2}t/2)$ and its Galilean gauge transformation (\ref{eq:gauge_NLSE}) \cite{Salman_2013} 
\[ \psi(s,t) \;=\; k_{0}\,\exp\left[i\tau_{0}\,\left( s - D\,\tau_{0}\,t\right) + iD\,k_{0}^{2}\,t/2 \right], \]
where $\tau_{0} \equiv v/(2D)$. Another simple separable solution of the NLSE (\ref{eq:NLSE}) is 
\begin{equation}
    \psi(s,t) \;=\; 2\,k\;{\rm sech}\left(k s\right)\;\exp(i\,D\,k^{2}\,t),
    \label{eq:optical_soliton}
\end{equation}
where the wavenumber $k$ is constant. This solution arises in the context of optical soliton propagation \cite{Hasegawa_2000,Hasegawa_2022}. We will return to this solution in Sec.~\ref{sec:Lame}, when we consider the periodic solution 
\begin{equation} 
\psi(s,t) \;\equiv\; \Psi(s)\,\exp(iD\,\epsilon^{2}k_{0}^{2}t/2),
\label{eq:Lame_sol}
\end{equation}
where $\Psi(s)$ is a periodic function of $s$ and $\epsilon$ is an eigenvalue parameter.

\subsubsection{Curvature envelope soliton}

Next, using our notation, Zakharov and Shabat \cite{ZS_1972} proposed the following traveling-wave envelope-soliton solution for the NLSE (\ref{eq:NLSE})
\begin{equation}
    \psi(s,t) \;=\; 2k\,{\rm sech}[k\,(s - c\,t)] \exp\left[i\,\Phi(s,t) \right],
    \label{eq:ZS_soliton}
\end{equation}
where the phase $\Phi(s,t)$ is defined, up to a constant, as
\begin{equation} 
\Phi(s,t) \equiv\; \frac{c}{2D}\left(s \;-\; \frac{c}{2}\,t\right) \;+\; Dk^{2}\,t,
\label{eq:eikonal_ZS}
\end{equation}
which is just the Galilean gauge transformation (\ref{eq:gauge_NLSE}) of Eq.~(\ref{eq:optical_soliton}).

In fact, this solution can be obtained by inserting the ansatz $\psi(s,t) \equiv \Psi(\xi)\,\exp(ik\,s - i\omega\,t)$ into the NLSE (\ref{eq:NLSE}), where $\xi \equiv k_{0}\,(s - ct)/2$ and the wavefunction $\Psi(\xi)$ is assumed to be a real-valued function. The imaginary part of the resulting equation yields $k = c/(2D)$, while the real part yields the second-order differential equation for the envelope function $\Psi(\xi)$:
\begin{equation}
    \Psi^{\prime\prime} \;=\; -\,\frac{2\,\Psi^{3}}{k_{0}^{2}} \;+\; \frac{4}{k_{0}^{2}} \left(k^{2} - \frac{\omega}{D}\right)\;\Psi.
\end{equation}
This equation can be solved in terms of Jacobi elliptic functions as $\Psi(\xi) \equiv k_{0}\,{\rm dn}(\xi|m)$, while the dispersion relation 
\begin{equation}
    \omega(k;m,k_{0}) \;\equiv\; D\,\left[k^{2} - k_{0}^{2}\,(2 - m)/4\right]    
\end{equation}
yields the wave group velocity $\partial\omega/\partial k = 2D\,k \equiv c$, as expected \cite{Zakharov_1968,Dewar_1972}. 

If we now substitute the dispersion relation into our ansatz, we obtain the self-modulation solution
\begin{equation}
    \psi(s,t) \;=\; k_{0}\;{\rm dn}[k_{0}\,(s - ct)/2\,|\,m]\;\exp[i\,\Theta(s,t;m)],
    \label{eq:NLSE_dn}
\end{equation}
where the eikonal phase $\Theta(s,t;k,\omega) \equiv k\,s - \omega\,t$ is 
\begin{equation} 
\Theta(s,t;m) \;=\; \frac{c\,s}{2D} \;+\; \frac{D}{4}\,\left[ k_{0}^{2}\,(2 - m) - \frac{c^{2}}{D^{2}}\right]\,t.
\label{eq:eikonal_dn}
\end{equation}
We now note that, in the limit $m \rightarrow 1$, we recover the Zakharov-Shabat envelope-soliton solution (\ref{eq:ZS_soliton}) from Eq.~(\ref{eq:NLSE_dn}).

\subsubsection{Torsion envelope soliton}

Hasimoto \cite{Hasimoto_1972}, on the other hand, considered the case of a spatial curve with constant torsion $\tau_{0}$, and obtained the NLSE solitary wave
\begin{equation}
    \psi(s,t) \;=\; 2\,\tau_{0}\;{\rm sech}\left[\tau_{0}\,(s - c\,t)\right]\;\exp(i\tau_{0}\,s),
    \label{eq:Hasimoto_soliton}
\end{equation}
where the constant wave speed is $c \equiv 2D\tau_{0}$. From the Frenet-Serret curvature $\kappa(s) = 2\,\tau_{0}\,{\rm sech}(\tau_{0}s)$ and the constant torsion $\tau(s) = \tau_{0}$, Hasimoto \cite{Hasimoto_1972} was then able to construct a spatial curve ${\bf r}(s)$. 

\subsection{\label{sec:breather}Breather soliton solutions}

An important class of soliton solutions of the NLSE (\ref{eq:NLSE}) are called {\it breather} solitons \cite{Dysthe_1999,Chabchoub_2014}, which are either periodic in time and localized in space, or periodic in space and localized in time. 

\begin{figure}
\epsfysize=2.2in
\epsfbox{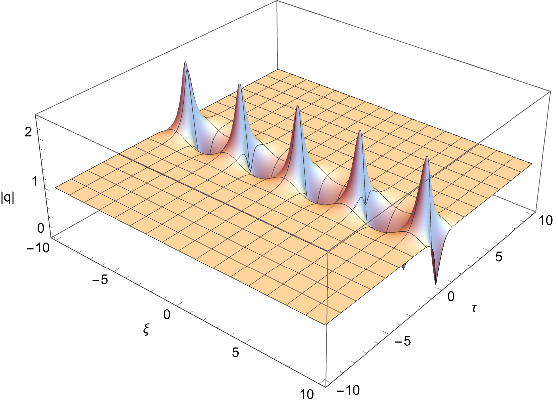}
\caption{3D Plot of the magnitude $|q_{A}(\xi,\tau;\pi/4)|$ for the Akhmediev breather soliton solution (\ref{eq:A_breather}), which is periodic in space $\xi$ and localized in time $\tau$.}
\label{fig:breather}
\end{figure}

First, to derive these solutions, we introduce the parametrization $\Omega(\phi) \equiv \sin(2\phi)$ and $p(\phi) = \sqrt{2}\,\sin\phi$ (where $0\leq \phi \leq \pi/2$), and we replace the time $t$ with the dimensionless time $\tau$ (Note: torsion does not appear in Sec.~\ref{sec:breather}), where $\omega\,t \equiv \Omega(\phi)\,\tau$, and replace the arclength distance $s$ with the dimensionless variable $\xi$, where $k\,s \equiv p(\phi)\,\xi$. Second, we define $\psi(s,t) \equiv A\,q(\xi,\tau)$ and, using the definitions $A = 2\,k/p(\phi)$ and $\omega = D\,k^{2}\Omega(\phi)/p^{2}(\phi)$, the NLSE (\ref{eq:NLSE}) becomes the dimensionless NLSE
\begin{equation}
    i\;\pd{q}{\tau} \;+\; \frac{\partial^{2}q}{\partial\xi^{2}} \;+\; 2\,|q|^{2}\,q \;=\; 0,
\end{equation}
whose solution is given by Dysthe and Trulsen \cite{Dysthe_1999} as the Akhmediev breather solution \cite{Akhmediev_1986,Akhmediev_2009a,Akhmediev_2009b}
\begin{equation}
    q_{A}(\xi,\tau;\phi) = \frac{-\,e^{i\tau}}{\sqrt{2}}\left[ \frac{\cosh(\Omega\tau - 2i\phi) - \cos\phi\,\cos(p\xi)}{\cosh(\Omega\tau) - \cos\phi\,\cos(p\xi)}\right],
    \label{eq:A_breather}
\end{equation}
which is periodic in space $\xi$ (with period $2\pi/p$) and is localized in time $\tau$, with the magnitude $|q_{A}(\xi,\tau;\phi)|$ reaching the maximum $1 + 2\,\cos\phi$ (Fig.~\ref{fig:breather} shows the magnitude $|q_{A}|$ of the Akhmediev breather for $\phi = \pi/4$). The Ma breather soliton solution \cite{Ma_1979}, on the other hand, which is periodic in time $\tau$ and localized in space $\xi$, can easily be expressed as \cite{Dysthe_1999}
\begin{equation}
    q_{M}(\xi,\tau;\phi) = \frac{-\,e^{i\tau}}{\sqrt{2}}\left[ \frac{\cosh\phi\,\cosh(p\xi) - \cos(\Omega\tau - 2i\phi)}{\cosh\phi\,\cosh(p\xi) - \cos(\Omega\tau)}\right],
    \label{eq:M_breather}
\end{equation}
where $\Omega = \sinh(2\phi)$ and $p = \sqrt{2}\,\sinh\phi$. We note that these breather solutions can be made to propagate with a dimensionless velocity $u$ along the $\xi$-axis by using the Galilean gauge transformation $(\xi,\tau) \rightarrow (\xi - u\,\tau,\tau)$ and $\exp(i\tau) \rightarrow \exp[i\tau + i(u\,\xi/2 - u^{2}\,\tau/4)]$.

\begin{figure}
\epsfysize=2.4in
\epsfbox{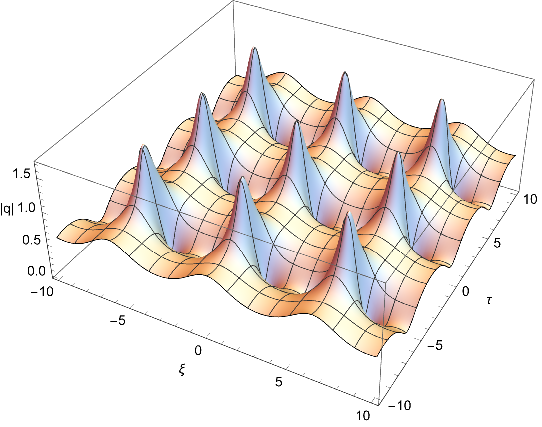}
\caption{3D Plot of the magnitude $|q(\xi,\tau;\mu = \pi/32)|$ for the lattice soliton solution (\ref{eq:q_Ap}), which is doubly periodic in space $\xi$ and in time $\tau$.}
\label{fig:lattice}
\end{figure}

\begin{widetext}
Finally, Akhmediev and Korneev \cite{Akhmediev_1986} showed that the spatially-periodic Akhmediev breather solution (\ref{eq:A_breather}) can be extended to a doubly-periodic lattice soliton solution
\begin{equation}
    q(\xi,\tau;\mu) = \frac{\cos(2\mu)}{\sqrt{2}} \left[ \frac{{\rm cd}(\cos\mu\,\xi|\tan^{2}\mu)\,{\rm cn}(\tau|\cos^{2}(2\mu)) + i\,\sqrt{2}\cos\mu\,{\rm sn}(\tau|\cos^{2}(2\mu))}{\sqrt{2}\,\cos\mu - {\rm cd}(\cos\mu\,\xi|\tan^{2}\mu)\,{\rm dn}(\tau|\cos^{2}(2\mu))}\right] e^{i\tau},
    \label{eq:q_Ap}
\end{equation}
where we introduced the modulus $k = \cos(2\mu)$, with $0 \leq \mu \leq \pi/4$, so that $(1 + k)/2 = \cos^{2}\mu$ and $(1 - k)/(1 + k) = \tan^{2}\mu$. We can easily check that, by using the standard limits of the Jacobi elliptic functions as $k \rightarrow 1$ \cite{NIST_Chap22}, we find $q(\xi,\tau;0) \equiv q_{A}(\xi,\tau;\pi/4)$, with $\Omega = 1 = p$ used in Eq.~(\ref{eq:A_breather}). Figure \ref{fig:lattice} shows the lattice soliton solution (\ref{eq:q_Ap}) for $\mu = \pi/32$, which exhibits the double periodicity in space $\xi$ and in time $\tau$. In the limit $\mu \rightarrow \pi/4$ (i.e., $k \rightarrow 0$), on the other hand, the doubly-periodic solution (\ref{eq:q_Ap}) becomes
\begin{eqnarray}
    \lim_{\mu \rightarrow \pi/4}q(\xi,\tau;\mu) &=& \frac{e^{2i\tau}}{\sqrt{2}} \lim_{\mu \rightarrow \pi/4}\left[ \frac{\cos(2\mu)}{\sqrt{2}\cos\mu - {\rm cd}(\cos\mu\,\xi|\tan^{2}\mu)\,{\rm dn}(\tau|\cos^{2}(2\mu))} \right] \nonumber \\
     &=& \frac{e^{2i\tau}}{\sqrt{2}} \left[ \frac{2}{1 + 2\,{\rm sinh}^{2}(\xi/\sqrt{2})}\right] = \sqrt{2}\,{\rm sech}(\sqrt{2}\xi)\;e^{2i\tau},
\end{eqnarray}
\end{widetext}
where we obtained the final result by applying L'H\^{o}pital's rule. We note that this solution is identical to the optical-soliton solution (\ref{eq:optical_soliton}), where $ks = \xi$ and $\omega t = \tau$, after performing a scaling transformation (\ref{eq:NLSE_scale}) with $a = \sqrt{2}$.

\section{\label{sec:sec_4}Nonlinear Schr\"{o}dinger Equation on an Elastica Knot}

Since the NLSE solutions discussed in Sec.~\ref{sec:simple_NLSE} are not connected to an elastica curve, we will not discuss these solutions further. In this Section, we show that the traveling-wave solution of the NLSE (\ref{eq:NLSE}) matches the elastica-knot curvature equation (\ref{eq:kappa_pp}) provided the elastica-knot constant $k_{0}^{2}\lambda = -\,c^{2}/(2D^{2})$ is expressed in terms of the traveling-wave NLSE parameters $(c,D)$. Next, we note that none of the envelope soliton solutions (\ref{eq:ZS_soliton}), (\ref{eq:NLSE_dn}), and (\ref{eq:Hasimoto_soliton}) are relevant here since their phases are not of the traveling-wave form. Lastly, using the quantum-mechanical definition $D \equiv \hbar/(2M)$ for the circulation parameter $D$, we find that the WKB representation of the wavefunction $\psi(s,t)$ is an exact solution of the NLSE (\ref{eq:NLSE}).

\subsection{Traveling-wave NLSE on an elastica knot}

We now seek a full traveling-wave solution $\psi(s,t) \equiv \Psi(s_{t})$ of the NLSE (\ref{eq:NLSE}), where $s_{t} \equiv s - c\,t$ is defined in terms of a constant velocity $c$. With this transformation, where $\partial_{t}\psi = -c\,\Psi^{\prime}$ and $\partial_{ss}^{2}\psi = \Psi^{\prime\prime}$, the NLSE (\ref{eq:NLSE}) transforms into the second-order ordinary differential equation
\begin{equation}
    i\,(c/D)\;\Psi^{\prime} \;=\; \Psi^{\prime\prime} \;+\; \kappa^{2}\;\Psi/2, \label{eq:Psi_NLSE}
\end{equation}
where a prime now denotes a derivative with respect to the wave-frame position $s_{t}$ along the curve.

Next, we substitute the Hasimoto {\it ansatz} 
\begin{equation}
    \Psi(s_{t}) \;\equiv\; \kappa(s_{t})\;\exp[i\theta(s_{t})],
    \label{eq:ansatz}
\end{equation} 
so that Eq.~(\ref{eq:Psi_NLSE}) can be separated, respectively, into the real and imaginary parts:
\begin{eqnarray}
    0 &=& (c/D)\,\kappa\;\theta^{\prime} \;+\; \kappa^{\prime\prime} \;-\; \kappa\,\theta^{\prime 2} \;+\; \frac{1}{2}\,\kappa^{3}, 
    \label{eq:NLSE_Re} \\
    0 &=& -\,(c/D)\;\kappa\,\kappa^{\prime} \;+\; \left(\kappa^{2}\,\theta^{\prime}\right)^{\prime}. \label{eq:NLSE_Im} 
\end{eqnarray}
The imaginary part (\ref{eq:NLSE_Im}) yields the phase equation
\begin{equation}
    \theta^{\prime} \;=\; \frac{c}{2D} \;+\; \frac{k_{0}^{2}}{\kappa^{2}}\;\tau_{0},
    \label{eq:theta_prime_st}
\end{equation}
where we introduced the constants $(k_{0},\tau_{0})$ from the torsion conservation law (\ref{eq:tau_eq}). 

When we substitute the phase equation (\ref{eq:theta_prime_st}) into the real part (\ref{eq:NLSE_Re}), we obtain the traveling-wave NLSE curvature equation
\begin{eqnarray}
    \kappa^{\prime\prime} \;+\; \frac{1}{2}\,\kappa^{3} &=& -\;\frac{c}{D}\;\kappa\,\theta^{\prime} \;+\; \kappa\;\theta^{\prime 2} \;=\; -\;\frac{c^{2}}{4D^{2}}\;\kappa \;+\; \frac{k_{0}^{4}
    \tau_{0}^{2}}{\kappa^{3}}.
    \label{eq:kappa_pp_gamma}
\end{eqnarray}
By comparing this equation with the elastica curvature equation (\ref{eq:kappa_pp}), we find that the dimensionless elastica constant
\begin{equation}
    \lambda \;=\; -\,\frac{c^{2}}{2k_{0}^{2}D^{2}} \;\equiv\; -\;\frac{1}{2}\,\gamma^{2} \leq 0
    \label{eq:lambda_gamma}
\end{equation}
is expressed in terms of the traveling-wave NLSE parameters $(c,D)$ and the knot curvature scale $k_{0}$. Hence, while the constant $\lambda$ was arbitrary in classical elastica knot theory, represented by the elastica curvature equation (\ref{eq:kappa_pp}), it acquires an immediate meaning in the traveling-wave NLSE elastica knot theory, represented by the traveling-wave NLSE curvature equation (\ref{eq:kappa_pp_gamma}).

Finally, the traveling-wave solution (\ref{eq:ansatz}) is expressed as
\begin{equation}
\Psi(s_{t}) \;=\; \kappa(s_{t})\; \exp\left( \frac{c\,s_{t}}{2D} \;+\; \int_{0}^{s_{t}} \tau(s')\;ds'\right),
\label{eq:TW_NLSE}
\end{equation}
where the curvature $\kappa(s_{t})$ is a solution of the NLSE elastica curvature equation (\ref{eq:kappa_pp_gamma}), the torsion $\tau(s_{t}) \equiv k_{0}^{2}\tau_{0}/\kappa^{2}(s_{t})$ is defined through the torsion conservation law (\ref{eq:tau_eq}), and the phase is obtained by integrating Eq.~(\ref{eq:theta_prime_st}), with the linear term $c\,s_{t}/(2D)$ associated with a Galilean gauge transformation (\ref{eq:gauge}).

\subsection{Elastica hydrodynamics equations}

A complementary point of view of the NLSE (\ref{eq:NLSE}) follows a hydrodynamics formulation \cite{Betchov_1965} in terms of fluid density $\rho(s,t)$ and a fluid density $u(s,t)$. For this purpose, we follow Zakharov \cite{Zakharov_1968} and Hasimoto \cite{Hasimoto_1972}, and we consider the WKB-like solution of the NLSE (\ref{eq:NLSE}):
\begin{equation} 
\psi(s,t) \;\equiv\; \sqrt{\rho(s,t)}\,\exp[i\,\Phi(s,t)/2],
\label{eq:psi_hydro}
\end{equation}
where $\rho = \kappa^{2}$ and the phase integral
\begin{equation}
    \Phi(s,t) \;=\; D^{-1}\,\int_{0}^{s}u(s',t) ds' \;+\; \Phi_{0}(t)
\end{equation} 
is expressed in terms of the fluid speed \cite{Hasimoto_1972}
\begin{equation}
u(s,t) \;=\; D\,\partial_{s}\Phi(s,t) \;\equiv\; 2D\,\tau(s,t),
\label{eq:u_hydro}
\end{equation}
and the time-dependent phase $\Phi_{0}(t)$ will be solved from the elastica hydrodynamics equations. A similar discussion is found in Sec.~7 of the work by Dewar \cite{Dewar_1972}.

First, when we insert Eq.~(\ref{eq:psi_hydro}) into the NLSE (\ref{eq:NLSE}), the imaginary part yields the hydrodynamics continuity equation \cite{Zakharov_1968}
\begin{equation} 
\pd{\rho}{t} \;+\; \pd{}{s}(\rho\,u) \;=\; 0, 
\label{eq:continuity}
\end{equation}
while the real part of the NLSE (\ref{eq:NLSE}) yields the phase equation \cite{Zakharov_1968}
\begin{equation}
    \pd{\Phi}{t} \;+\; \frac{u^{2}}{2D} \;=\; D \left(\rho \;+\; 2\,{\cal B}\right),
\label{eq:phase}
\end{equation}
where ${\cal B} \equiv (\partial^{2}_{ss}\sqrt{\rho})/\sqrt{\rho}$ appears in the {\it Bohmian} quantum potential energy \cite{Bohm_1952}
\begin{equation}
    {\cal Q} \;=\; -\;\frac{\hbar^{2}}{2M}\;\left(\frac{1}{\sqrt{\rho}}\;\frac{\partial^{2}\sqrt{\rho}}{\partial s^{2}}\right) \;\equiv\; -\,\hbar\,D\,
    {\cal B}.
\end{equation}
Since ${\cal B} = \kappa^{\prime\prime}/\kappa$, we can use the NLSE elastica curvature equation (\ref{eq:kappa_pp_gamma}) to obtain
\begin{equation} 
\rho + 2\,{\cal B} \;\equiv\; \kappa^{2} \;+\; 2\,\kappa^{\prime\prime}/\kappa \;=\; -\,c^{2}/(2D^{2}) \;+\; 
 2\,k_{0}^{4}\tau_{0}^{2}/\kappa^{4},
 \label{eq:hydro}
\end{equation}
where the first term on the right is a constant.

Second, we take the partial derivative of the phase equation (\ref{eq:phase}) with respect to $s$, and using $\partial\Phi/\partial s \equiv u/D$, we obtain the fluid acceleration equation
\begin{eqnarray} 
\pd{u}{t} \;+\; u\,\pd{u}{s} &=& D^{2} \pd{}{s}\left(\rho \;+\; 2\,{\cal B}\right)
\;=\; -\;8D^{2}\,\frac{k_{0}^{4}\tau_{0}^{2}}{\kappa^{5}}\;\pd{\kappa}{s},
\label{eq:acceleration}
\end{eqnarray}
where we used Eq.~(\ref{eq:hydro}). When we combine this equation with the continuity equation (\ref{eq:continuity}), by multiplying it with $\rho = \kappa^{2}$, we obtain the hydrodynamic momentum equation
\begin{eqnarray}
    \pd{}{t}(\rho\,u) + \pd{}{s}\left(\rho\,u^{2}\right) &=& 4D^{2} \pd{}{s}\left(
    \frac{k_{0}^{4}\tau_{0}^{2}}{\kappa^{2}}\right).
    \label{eq:momentum}
\end{eqnarray}
We note that, for a general spatial curve with curvature $\kappa(s)$, Hasimoto \cite{Hasimoto_1972} uses the identity
\[ \rho\,\pd{}{s}(\rho + 2\,{\cal B}) \;\equiv\; \pd{}{s}\left(\frac{\rho^{2}}{2} + \rho\;\frac{\partial^{2}\ln\rho}{\partial s^{2}}\right), \]
 which appears as Eq.~(A7) in Hasimoto's work \cite{Hasimoto_1972}.

Third, we return to an elastica spatial curve and, by substituting $\rho = \kappa^{2}$ and $u \equiv 2D\,\tau$, as well as using the torsion conservation law (\ref{eq:tau_eq}), we find
\begin{equation}
\left. \begin{array}{lcl} 
\rho\,u &\equiv& 2D\,\kappa^{2}\tau \;=\; 2D\,k_{0}^{2}\tau_{0} \\ 
&& \\
\rho\,u^{2} &\equiv& 4D^{2}\,\kappa^{2}\tau^{2} \;=\; 4D^{2}k_{0}^{4}\tau_{0}^{2}/\kappa^{2}
\end{array} \right\}, 
\end{equation}
and, therefore, Eq.~(\ref{eq:momentum}) is trivially solved, since $\rho\,u$ is a constant and the two partial derivatives cancel out exactly. 

Finally, by substituting Eq.~(\ref{eq:hydro}) into the acceleration equation (\ref{eq:acceleration}), we obtain
\begin{equation}
\pd{u}{t} \;=\; 2D\,\pd{\tau}{t} \;=\; \pd{}{s}\left[ D^{2}\,(\rho + 2\,{\cal B}) \;-\; \frac{u^{2}}{2} \right] \;=\; 0,
\label{eq:u_t}
\end{equation}
which implies that the torsion $\tau(s)$ is a function of $s$ only. The continuity equation (\ref{eq:continuity}), on the other hand, implies that $\partial\kappa/\partial t = 0$, i.e., the curvature $\kappa(s)$ is also a function of $s$ only, which is consistent with the torsion conservation law (\ref{eq:tau_eq}). Next, by making use Eq.~(\ref{eq:u_t}), the phase equation (\ref{eq:phase}) becomes
\begin{equation}
\pd{\Phi}{t} \;=\; \frac{d\Phi_{0}(t)}{dt} \;=\; D\,(\rho + 2\,{\cal B}) \;-\; \frac{u^{2}}{2D} \;=\; -\;\frac{c^{2}}{2D},
\end{equation}
whose solution can be used to write the elastica hydrodynamic NLSE solution
\begin{equation}
\psi(s,t) \;=\; \kappa(s)\;\exp\left[ i\;\int_{0}^{s}\tau(s')\,ds' \;-\; \frac{i\,c^{2}t}{4D} \right].
\label{eq:psi_hydro_def}
\end{equation}
If we now apply the Galilean gauge transformation (\ref{eq:gauge}), we obtain
\begin{equation}
\psi(s,t) \;\rightarrow\; \kappa(s_{t})\;\exp\left[ i\;\int_{0}^{s_{t}}\tau(s')\,ds' \;-\; \frac{i\,c^{2}t}{4D} \;+\; \frac{i\,c}{2D} \left( s_{t} \;+\; \frac{ct}{2}\right) \right] \;=\; \kappa(s_{t})\;\exp\left[ i\;\int_{0}^{s_{t}}\tau(s')\,ds' \;+\; \frac{i\,c\,s_{t}}{2D}  \right],
\label{eq:TW_NLSE_hydro}
\end{equation}
which corresponds exactly to the traveling-wave elastica NLSE solution (\ref{eq:TW_NLSE}).

\section{\label{sec:sec_5}Elliptic solution of the elastica curvature equation}

In this Section, we present an explicit solution of the elastica curvature equation (\ref{eq:kappa_pp}) [or Eq.~(\ref{eq:kappa_pp_gamma})] in terms of the Jacobi elliptic functions and the Weierstrass elliptic functions. Tutorial presentations on the applications of elliptic functions in classical mechanics are presented in Refs.~\cite{Brizard_2015,Brizard_2009}, while additional applications of elliptic functions in plasma physics can be found in a recent paper \cite{Brizard_2026}.

We now return to an elastica knot, and seek a solution of the NLSE elastica curvature equation (\ref{eq:kappa_pp_gamma}), which is expressed as
\begin{equation}
    \kappa^{\prime\prime}(s) \;=\; -\;\frac{1}{2}\,\left[\kappa^{3}(s) \;+\; \frac{1}{2}\;k_{0}^{2}\gamma^{2}\kappa(s) \right] \;+\; \frac{k_{0}^{6}\nu^{2}}{4\,\kappa^{3}(s)},
    \label{eq:kappa_elastica}
\end{equation}
where we introduced the dimensionless torsion constant $\nu \equiv 2\,\tau_{0}/k_{0}$ and we substituted the traveling-wave parameter (\ref{eq:lambda_gamma}). 

First, we multiply this equation $\kappa^{\prime}$ and integrate, using the initial conditions $\kappa(0) = k_{0}$ and $\kappa^{\prime}(0) = 0$, to obtain
\begin{eqnarray}
    (\kappa^{\prime})^{2} &=& \frac{1}{4} \left( k_{0}^{4} - \kappa^{4}\right) \;+\; \frac{k_{0}^{2}\gamma^{2}}{4}\,\left(k_{0}^{2} - \kappa^{2}\right) \;+\; \frac{1}{4}\,k_{0}^{4}\nu^{2}\,\left(1 \;-\; \frac{k_{0}^{2}}{\kappa^{2}}\right).
    \label{eq:kappa_p2}
\end{eqnarray}
Next, we multiply this equation by $4\,\kappa^{2}$ to obtain
\begin{equation}
    [(\kappa^{2})^{\prime}]^{2} = -\;\kappa^{6} - k_{0}^{2}\gamma^{2}\,\kappa^{4} + k_{0}^{4}\kappa^{2}\,(1 + \nu^{2} + \gamma^{2}) - k_{0}^{6}\,\nu^{2}.
    \label{eq:kappa2_eq}
\end{equation}
The classical Jacobi solution of the elastica curvature equation (\ref{eq:kappa2_eq}) is given by Langer and Singer \cite{Langer_Singer_1984} as
\begin{equation}
    \kappa^{2}(s) \;=\; k_{0}^{2} \left[ 1 \;-\; \frac{p^{2}}{w^{2}}\,
    {\rm sn}^{2}(rs,p) \right],
    \label{eq:LS_kappa}
\end{equation}
where $r = k_{0}/(2w)$ is a constant parameter and ${\rm sn}(x,p)$ denotes a Jacobi function \cite{Lawden,NIST_Chap22} with a real argument $x = rs$ and a real modulus $p$, which is assumed to lie in the classical range $0 \leq p^{2} \leq w^{2} \leq 1$, which corresponds to a right triangle in parameter space $(p^{2},w^{2})$, with vertices at $(0,0)$, $(0,1)$, and $(1,1)$.

\subsection{Jacobi elliptic solution}

In the present work, we follow a slightly different approach, where we insert $\kappa^{2}(s) \equiv k_{0}^{2}\,q(v)/q_{0}$ into Eq.~(\ref{eq:kappa2_eq}), where $v \equiv \xi/\sqrt{q_{0}} = (\frac{1}{2}k_{0}s)/\sqrt{q_{0}}$ is a dimensionless variable, and $q_{0} \equiv q(0)$ is a constant. Hence, we obtain the dimensionless equation
\begin{eqnarray}
    \left(q^{\prime}\right)^{2} &=& -\,4q^{3} - 4\gamma^{2}\,q_{0}\,q^{2} + 4(1 + \nu^{2} + \gamma^{2})\,q_{0}^{2}\,q - 
    4\,q_{0}^{3}\nu^{2} \nonumber \\
    &=& 4\,(q_{0} - q)\;\left[ q^{2} \;+\; (1 + \gamma^{2})\,q_{0}q \;-\; q_{0}^{2}\nu^{2}\right],
    \label{eq:q_prime2}
\end{eqnarray}
which satisfies the initial condition $q^{\prime}(0) = 0$. 

Next, we consider the ansatz $q(v) \equiv q_{0} - a\,P^{2}(v)$, where $a$ is a constant, so that Eq.~(\ref{eq:q_prime2}) becomes
\begin{equation}
    \left(P^{\prime}\right)^{2} \;=\; a\,P^{4} \;-\; (3 + \gamma^{2})\,q_{0}\;P^{2} \;+\; (2 + \gamma^{2} - \nu^{2})\,
    q_{0}^{2}/a,
    \label{eq:P_Jacobi}
\end{equation}
which is to be solved under the initial conditions $P(0) = 0$ and $P^{\prime}(0) = 1$. Hence, these initial conditions require that $a = q_{0}^{2}\,(2 + \gamma^{2} - \nu^{2})$. 

Here, using the conventional notation \cite{AS}, three choices arise for the Jacobi elliptic function $P(v)$: (i) ${\rm sn}(v|m)$, (ii) ${\rm sc}(v|m)$, or (iii) ${\rm sd}(v|m)$. For case (i), we find $a = m$ and $-(3 + \gamma^{2})\,q_{0} = -\,(1 + m)$; for case (ii), we find $a = 1- m$ and $-(3 + \gamma^{2})\,q_{0} = 2- m$; and for case (iii), we find $a = -m\,(1-m)$ and $-(3 + \gamma^{2})\,q_{0} = 2m - 1$.

\subsubsection{Jacobi elastica solution in $(m,q_{0})$ space}

If we assume that $(m,q_{0})$ are both positive, we select case (i) for the Jacobi elliptic solution for $q(v)$:
\begin{equation}
    q(v) \;\equiv\; q_{0} \;-\; m\;{\rm sn}^{2}(v|m),
\end{equation}
so that the solution for the curvature differential equation (\ref{eq:kappa_elastica}) is expressed as \cite{LS_1984,Langer_Singer_1996,Barros_2018}
\begin{equation}
    \kappa^{2}(s) \;=\; k_{0}^{2} \left[ 1 \;-\; \frac{m}{q_{0}}\;
    {\rm sn}^{2}\left(\left.\frac{k_{0}s}{2\sqrt{q_{0}}}\right|m\right) \right],
    \label{eq:elastica_J}
\end{equation}
which exactly corresponds to the Langer-Singer solution (\ref{eq:LS_kappa}), with $p^{2} = m$, and $w^{2} = q_{0}$, with $rs = \xi/\sqrt{q_{0}}$. Here, the parameter $m$ in Eq.~(\ref{eq:elastica_J}) is simultaneously defined as
\begin{equation}
m \;=\; q_{0}\,(3 + \gamma^{2}) \;-\; 1 \;=\; q_{0}^{2}\,\left(2 + \gamma^{2} - \nu^{2}\right),
\label{eq:m_extended}
\end{equation}
which is not positive definite and, therefore, allows for an extension to the range $m < 0$ not considered (and in fact forbidden) in the standard theory of elastica knots \cite{Langer_Singer_1984,LS_1984}. From these relations, we obtain the following explicit functions 
\begin{eqnarray}
    \gamma^{2}(m,q_{0}) &=& \frac{(1 + m)}{q_{0}} - 3, \label{eq:lambda_mq0} \\
    \nu^{2}(m,q_{0}) &=& \frac{(1 - q_{0})\,(q_{0} - m)}{q_{0}^{2}}. \label{eq:nu2_mq0}
\end{eqnarray} 
We note that, according to Eq.~(\ref{eq:lambda_mq0}), the wave speed $c(m,q_{0}) = k_{0}D\,\gamma(m,q_{0})$ of the traveling-wave NLSE solution (\ref{eq:ansatz}), and the torsion parameter $\tau_{0}(m,q_{0}) = k_{0}\nu(m,q_{0})/2$, both depend on the knot parameters $(m,q_{0})$.

\subsubsection{NLSE elastica knot parameter space}

\begin{figure}
\epsfysize=1.7in
\epsfbox{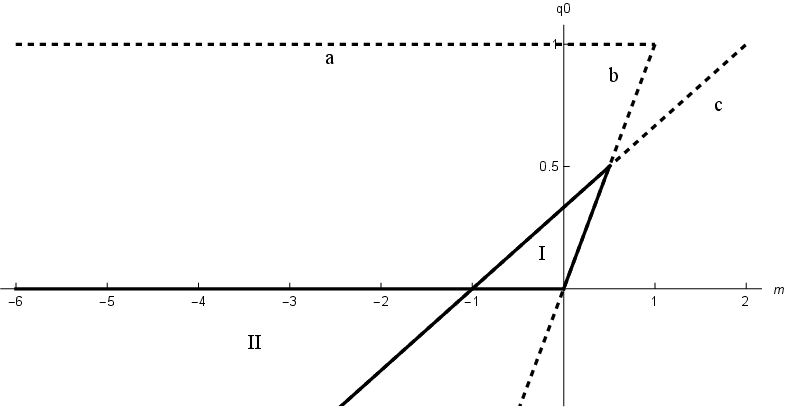}
\caption{Plots of the curves $\nu^{2} = 0$ (a: $q_{0} = 1$ and b: $q_{0} = m$) and the curve $\gamma^{2} = 0$ (c: $q_{0} = (1+m)/3$) in the $(m,q_{0})$ plane. We note that both $\nu^{2} > 0$ and $\gamma^{2} > 0$ inside the closed triangle I and the open triangle II (where $q_{0} < 0$ and $m < -1$).}
\label{fig:gammanu_2}
\end{figure}

Figure \ref{fig:gammanu_2} shows the curves a: $q_{0} = 1$ and b: $q_{0} = m$ associated with $\nu^{2} = 0$, and the curve c: $q_{0} = (1 + m)/3$ associated with $\gamma^{2} = 0$. Both functions (\ref{eq:lambda_mq0}) and (\ref{eq:nu2_mq0}) are positive either inside region I, defined as the closed triangle ($0 < q_{0} < \frac{1}{2}$ and $-1 < m < \frac{1}{2}$), or inside region II, defined as the open triangle ($m < -1$ and $(1 + m) < 3\,q_{0} < 0$). While the parameter space $(m,q_{0})$ for the classical elastica knot considered by Langer and Singer \cite{Langer_Singer_1984} lies inside the triangle $0 \leq m \leq q_{0} \leq 1$, our work greatly extends the parameter space by $m \leq q_{0} \leq 1$, where both $m$ and $q_{0}$ can be negative (e.g., region II).

\subsubsection{Extended Jacobi elliptic solution}

When both $q_{0}$ and $m$ are negative (region II of Fig.~\ref{fig:gammanu_2}), the argument $z = k_{0}s/(2\sqrt{q_{0}}) \equiv i\,y$ becomes imaginary and, using the identity (\ref{eq:sn_y}), the solution (\ref{eq:elastica_J}) becomes
\begin{equation}
    \kappa^{2}(s) \;=\; k_{0}^{2} \left[ 1 \;+\; \frac{n}{|q_{0}|}\;{\rm sd}^{2}\left(\left.\frac{k_{0}s}{2\sqrt{|q_{0}|n'}}\right|n'\right) \right],
    \label{eq:elastica_Jext}
\end{equation}
which is periodic, with a real period $k_{0}S(m,q_{0})/2 = 2\sqrt{|q_{0}|n'}\;{\sf K}(n^{\prime})$, where $n'(m) = 1/(1-m)$. We note that, while the classical solution (\ref{eq:elastica_J}) has a maximum $k_{0}^{2}$ at $s = 0$, the extended solution (\ref{eq:elastica_Jext}) has a maximum $k_{0}^{2}\,(1 + 1/|q_{0}|) > k_{0}^{2}$ at the half-period $k_{0}s/2 = \sqrt{|q_{0}|n'}\,{\sf K}(n')$.

\subsection{Weierstrass elliptic solution}

An alternate solution of the elastica curvature equation (\ref{eq:kappa2_eq}) is expressed in terms of the Weierstrass elliptic function $\wp(z;{\sf g}_{2},{\sf g}_{3})$, which is a solution of the differential equation \cite{Lawden,Brizard_2009,Brizard_2026,NIST_Chap23,WW}
\begin{equation}
    \left[\wp^{\prime}(z)\right]^{2} \;=\; 4\,\wp^{3}(z) \;-\; {\sf g}_{2}\,\wp(z) \;-\; {\sf g}_{3},
    \label{eq:wp_eq}
\end{equation}
where the lattice invariants $({\sf g}_{2},{\sf g}_{3})$ are defined in terms of the cubic roots $({\sf e}_{1},{\sf e}_{2},{\sf e}_{3})$, which satisfy the identity ${\sf e}_{1} + {\sf e}_{2} + {\sf e}_{3} = 0$, as
\begin{equation}
\left. \begin{array}{rcl}
{\sf g}_{2} &\equiv& 2\,({\sf e}_{1}^{2} + {\sf e}_{2}^{2} + {\sf e}_{3}^{2}) \\
{\sf g}_{3} &\equiv& 4\,{\sf e}_{1}\,{\sf e}_{2}\,{\sf e}_{3}
\end{array} \right\}.
\label{eq:g23_def}
\end{equation}
The Weierstrass elliptic function is doubly periodic \cite{NIST_Chap23,WW}, with half periods $(\omega_{1},\omega_{2},\omega_{3})$ defined as $\wp(\omega_{k}) \equiv {\sf e}_{k}$ and $\wp(z + 2\omega_{k}) \equiv \wp(z)$, with $k = 1,2,3$.

The Weierstrass elliptic solution of the elastica curvature equation (\ref{eq:kappa2_eq}) is expressed as
 \begin{equation}
    \kappa^{2}(s) \;=\; k_{0}^{2} \left[ 1 \;+\; {\sf e}_{3} \;-\; \wp\left(\frac{k_{0}s}{2} + \omega_{3}\right) \right],
    \label{eq:W_kappa}
\end{equation}
where ${\sf e}_{3} \equiv \wp(\omega_{3})$ is used to satisfy the initial condition $\kappa^{2}(0) = k_{0}^{2}$. When Eq.~(\ref{eq:W_kappa}) is substituted into Eq.~(\ref{eq:kappa2_eq}), where ${\sf e}_{3} = -\gamma^{2}/3 - 1$ removes the quadratic term in Eq.~(\ref{eq:wp_eq}), and the lattice invariants $({\sf g}_{2},{\sf g}_{3})$ are defined as
\begin{eqnarray}
    {\sf g}_{2}(\nu,\gamma) &\equiv& 4 \left(1 \;+\; \nu^{2} \;+\; \gamma^{2} \;+\; \gamma^{4}/3\right), \label{eq:g2_gammanu} \\
    {\sf g}_{3}(\nu,\gamma) &\equiv& 4 \left[ \nu^{2} \;+\; \gamma^{2} \left(1 \;+\; \nu^{2} \;+\; \gamma^{2}\right)/3 \;+\; 2\,\gamma^{6}/27\right]. \label{eq:g3_gammanu}
\end{eqnarray}
Since these invariants, as well as the discriminant $\Delta \equiv {\sf g}_{2}^{3} - 27\,{\sf g}_{3}^{2}$, are positive definite, the half-period $\omega_{1}$ is real, the half-period $\omega_{3}$ is imaginary, and the half-period $\omega_{2} \equiv -\omega_{1} - \omega_{3}$ is complex valued.

\subsubsection{Weierstrass elastica solution in $(m,q_{0})$ space}

Based on mathematical grounds \cite{WW}, see Ref.~\cite{Brizard_2009} for explicit examples involves classical mechanics, there is a connection between the Weierstrass solution (\ref{eq:W_kappa}) and the Jacobi solution (\ref{eq:elastica_J}). For this purpose, we use the identity (see formula 23.6.26 of Ref.~\cite{NIST_Chap23})
\begin{equation}
    m\,\left(\frac{{\sf K}(m)}{\omega_{1}}\right)^{2}\;{\rm sn}^{2}\left(\left.\frac{{\sf K}(m)\xi}{\omega_{1}}\right|m\right) \;\equiv\; \wp(\xi + \omega_{3}) - {\sf e}_{3},
    \label{eq:JW_id}
\end{equation}
where $\xi = k_{0}s/2$ and $m \equiv ({\sf e}_{2} - {\sf e}_{3})/({\sf e}_{1} - {\sf e}_{3})$ is defined in terms of the Weierstrass cubic roots $({\sf e}_{1},
{\sf e}_{2},{\sf e}_{3})$. By requiring that \cite{NIST_Chap23}
\begin{equation} 
{\sf K}(m)/\omega_{1} \;=\; \sqrt{{\sf e}_{1} - {\sf e}_{3}} \;\equiv\; 1/\sqrt{q_{0}},
\label{eq:K_q0}
\end{equation}
the identity (\ref{eq:JW_id}) becomes
\begin{equation}
    (m/q_{0})\;{\rm sn}^{2}(\xi/\sqrt{q_{0}}|m) \;\equiv\; \wp(\xi + \omega_{3}) \;-\; {\sf e}_{3},
    \label{eq:sn_P}
\end{equation}
and the cubic roots are \cite{NIST_Chap23}
\begin{eqnarray}
    {\sf e}_{1} &=& (2 - m)/(3q_{0}), \\
    {\sf e}_{2} &=& (2m - 1)/(3q_{0}), \\
    {\sf e}_{3} &=& -\;(1 + m)/(3q_{0}),
\end{eqnarray}
which satisfy the constraint ${\sf e}_{1} + {\sf e}_{2} + {\sf e}_{3} = 0$. Associated with these cubic roots are the Weierstrass lattice invariants (\ref{eq:g23_def}), and the discriminant $\Delta$, which are expressed as
\begin{eqnarray}
    {\sf g}_{2} &=& \frac{4}{3q_{0}^{2}}\,\left(1 - m + m^{2}\right), \label{eq:g2_def} \\
    {\sf g}_{3} &=& \frac{4}{27\,q_{0}^{3}}\,\left( 2 - 3m - 3 m^{2} + 2m^{3}\right), \label{eq:g3_def} \\
    \Delta &=& 16\,m^{2}\,(1 - m)^{2}/q_{0}^{6}.
\end{eqnarray}
Here, we note that, ${\sf g}_{2} > 0$ for all real values of $(m,q_{0})$, while ${\sf g}_{3} > 0$ in the range $-1 < m < \frac{1}{2}$ when $q_{0} > 0$ (i.e., region I in Fig.~\ref{fig:gammanu_2}) or in the range $m < m_{0}^{-} \simeq -4.7510$ and $q_{0} < 0$ (i.e., region II in Fig.~\ref{fig:gammanu_2}), while $g_{3} < 0$ in the range $m_{0}^{-} < m < -1$ or $\frac{1}{2} < m < 1$. Hence, the Weierstrass invariants $({\sf g}_{2},{\sf g}_{3},\Delta)$ are positive in the two regions I and II associated with the constraints $\nu^{2} > 0$ and $\gamma^{2} > 0$.

Next, we introduce the new parametric function $v(m,q_{0})$ through the identity 
\begin{equation}
1 \;\equiv\; (m/q_{0})\,{\rm sn}^{2}(v|m) \;=\; \wp(v + \omega_{3}) - {\sf e}_{3},  
\label{eq:1_v}
\end{equation}
where 
\begin{equation} 
    v(m,q_{0}) \;\equiv\; \sqrt{q_{0}}\;F(\phi|m) \;=\; \int_{0}^{\phi}\frac{\sqrt{q_{0}}\;d\theta}{\sqrt{1 - m\,\sin^{2}\theta}}
    \label{eq:v_Fphi}
\end{equation}
is defined in terms of the incomplete elliptic integral of the first kind $F(\phi|m)$, with $\phi(m,q_{0}) \equiv \arcsin(\sqrt{q_{0}/m})$. We also note that $[\wp^{\prime}(v + \omega_{3})]^{2} = -\,4\,\nu^{2}$, which yields the definition
\begin{equation}
    \wp^{\prime}(v + \omega_{3}) \;\equiv\; 2i\,\nu.
    \label{eq:wp_v}
\end{equation} 
Hence, the Weierstrass elliptic solution of the elastica-knot curvature equation is given by Eq.~(\ref{eq:W_kappa}), which is periodic with period $k_{0}S/2 = 2\,\omega_{1} \equiv 2\sqrt{q_{0}}\,{\sf K}(m)$. This solution, not explored in the past literature, is entirely equivalent to the Jacobi elliptic solution (\ref{eq:elastica_J}), and will be used in Secs.~\ref{sec:angle} and \ref{sec:TW_NLSE}.

 \section{\label{sec:sec_6}Elastica-knot Spatial Curve}

 In this Section, we will construct a spatial curve corresponding to a closed elastica knot, which is defined in terms of the Frenet-Serret curvature $\kappa(s)$ and torsion $\tau(s)$. Here, the curvature $\kappa(s)$ is assumed to be a solution of the traveling-wave NLSE curvature equation (\ref{eq:kappa_pp_gamma}) while the torsion $\tau(s)$ satisfies the conservation law $\tau(s) = \tau_{0}\,k_{0}^{2}/\kappa^{2}(s)$.
 
 Since the vector ${\bf W}$, defined by Eq.~(\ref{eq:W_def}), is a constant along the spatial curve ${\bf r}(s)$, it is natural to choose a cylindrical representation $(\rho,\varphi,z)$, where the vertical unit vector is defined as
 \begin{equation}
     \wh{\sf z} \;\equiv\; \frac{\bf W}{|{\bf W}|} \;=\; A(s)\,
     \wh{\sf t} \;+\; B(s)\,\wh{\sf n} \;+\; C(s)\,\wh{\sf b}. 
     \label{eq:z_FS}
 \end{equation}
Here, Eq.~(\ref{eq:W_const}) yields the coefficients
 \begin{eqnarray}
     A(s) &=& {\cal R}^{2}\,\left(\kappa^{2}(s) \;+\; \gamma^{2}
     k_{0}^{2}/2\right)/2, \label{eq:A_def} \\
     B(s) &=& {\cal R}^{2}\;\kappa^{\prime}(s), \label{eq:B_def} \\
     C(s) &=& {\cal R}^{2}\,k_{0}^{2}\tau_{0}/\kappa(s), \label{eq:C_def}
 \end{eqnarray}
which satisfy the constraint $A^{2} + B^{2} + C^{2} = 1$ derived from Eq.~(\ref{eq:kappa_p2}). Using Eqs.~(\ref{eq:W_const}) and (\ref{eq:kappa_p2}), the constant magnitude of 
${\bf W}$ is defined as
 \begin{equation}
     |{\bf W}|^{2} \;=\; \frac{k_{0}^{4}}{4} \left[ \left(1 + \frac{\gamma^{2}}{2}\right)^{2} \;+\; \nu^{2}\right] \;\equiv\; {\cal R}^{-4},
 \end{equation}
 where $k_{0}{\cal R}(m,q_{0})$ depends on the parameters $(m,q_{0})$.
 
Next, the cylindrical representation of the spatial curve is expressed as
 \begin{equation}
     {\bf r}(s) \;=\;\rho(s)\left[ \cos\varphi(s)\,\wh{\sf x} + \sin\varphi(s)\,\wh{\sf y}\right] \;+\; z(s)\,\wh{\sf z} \;\equiv\; \rho(s)\,\wh{\rho}(s) \;+\; z(s)\,\wh{\sf z}.
     \label{eq:r_cyl}
 \end{equation}
In the present work, we are interested in spatial curves (\ref{eq:r_cyl}) that are {\bf closed}, which are defined by the following two conditions: (i) the radial and vertical positions $\rho(s)$ and $z(s)$ must be periodic functions of $s$; and (ii) the azimuthal angular shift $\Delta\varphi$ over a spatial period must be a rational fraction of $\pi$. We note that, if only the condition (i) is satisfied, the resulting spatial curve is known as a torus knot \cite{Langer_Singer_1984} . In what follows, we will show that the solutions for the cylindrical coordinates $(\rho,\varphi,z)$ are expressed as functions of $s$ while being parametrized by the elastica parameters $(m,q_{0})$. While only the radial position $\rho(s)$ will be intrinsically periodic [see Eq.~(\ref{eq:rho_sol})], the periodicity of the vertical position $z(s)$ will introduce a constraint $q_{0} \equiv Q(m)$ in the $(m,q_{0})$ parameter space, which must be extended to the range $m < 0$ in order to obtain NLSE traveling-wave solutions along closed elastica knots.

 \subsection{Frenet-Serret triad in cylindrical geometry}

With the vertical unit vector $\wh{\sf z}$ given by Eq.~(\ref{eq:z_FS}), we now derive expressions for the radial unit vector $\wh{\rho}(\varphi)$ and the azimuthal unit vector $\wh{\varphi}(\varphi)\equiv \partial\wh{\rho}(\varphi)/\partial\varphi = \wh{\sf z}\btimes\wh{\rho}$. Since the unit vectors 
$(\wh{\rho},\wh{\varphi})$ are perpendicular to ${\bf W} = {\cal R}^{-2}\,\wh{\sf z}$, we are free to choose
\begin{equation}
    \wh{\rho} \;\equiv\; \frac{\wh{\sf z}\btimes\wh{\sf b}}{|\wh{\sf z}\btimes\wh{\sf b}|} \;=\; \frac{-\,A\,\wh{\sf n} + B\,\wh{\sf t}}{\sqrt{1 - C^{2}}},
    \label{eq:rho_FS}
\end{equation}
where we used $A^{2} + B^{2} = 1 - C^{2}$, and
\begin{equation}
    \wh{\varphi} \;\equiv\; \wh{\sf z}\btimes\wh{\rho} \;=\;
    \frac{C\,\wh{\sf z} \;-\; \wh{\sf b}}{\sqrt{1 - C^{2}}}.
    \label{eq:varphi_FS}
\end{equation}
Hence, using Eqs.~(\ref{eq:z_FS}) and (\ref{eq:rho_FS})-(\ref{eq:varphi_FS}), the cylindrical representation of the tangent unit vector
\begin{equation}
    \wh{\sf t}(s) \equiv {\bf r}^{\prime} =  
    \rho^{\prime}(s) \,\wh{\rho}(s) \;+\; \rho(s)\,
    \varphi^{\prime}(s)\,\wh{\varphi}(s) \;+\; z^{\prime}(s)\,
    \wh{\sf z}
    \label{eq:t_cyl}
\end{equation}
yields the expressions
\begin{eqnarray}
    \rho^{\prime}(s) &\equiv& \wh{\sf t}\bdot\wh{\rho} \;=\; B(s)/\sqrt{1 - C^{2}(s)}, \label{eq:rho_prime} \\
    \rho(s)\,\varphi^{\prime}(s) &\equiv& \wh{\sf t}\bdot\wh{\varphi} \;=\; A(s)\,C(s)/\sqrt{1 - C^{2}(s)}, \label{eq:varphi_prime} \\
    z^{\prime}(s) &\equiv& \wh{\sf t}\bdot\wh{\sf z} \;=\; A(s). \label{eq:z_prime} 
\end{eqnarray}
Here, using the definition 
\begin{equation} 
\Omega^{2} \;\equiv\; \left(\frac{1}{2}\,{\cal R}^{2}k_{0}^{2}\,\nu\right)^{2} \;=\; \frac{4\,\nu^{2}}{[(2 + \gamma^{2})^{2} + 4\,\nu^{2}]} \;<\; 1,
\label{eq:Omega_def}
\end{equation}
we find $1 - C^{2}(s) = [\kappa^{2}(s) - k_{0}^{2}\,\Omega^{2}]/\kappa^{2}(s)$, and $1 - C^{2}(0) = 1 - \Omega^{2} > 0$.

Finally, we note that the Frenet-Serret triad $(\wh{\sf t},\wh{\sf n},\wh{\sf b})$ can be expressed in terms of the cylindrical frame $(\wh{\rho},\wh{\varphi},\wh{\sf z})$ by inverting the expressions (\ref{eq:z_FS}), (\ref{eq:rho_FS}), and (\ref{eq:varphi_FS}), whose inversion yields the Frenet-Serret unit vectors in cylindrical geometry 
\begin{eqnarray}
    \wh{\sf t} &=& \sin\alpha\;\wh{\rho} \;+\; \cos\alpha\;\left( \sin\beta\,
    \wh{\varphi} + \cos\beta\,\wh{\sf z}\right), \label{eq:t_ab} \\
    \wh{\sf n} &=& -\,\cos\alpha\;\wh{\rho} \;+\; \sin\alpha\;\left( \sin\beta\,
    \wh{\varphi} + \cos\beta\,\wh{\sf z}\right), \label{eq:n_ab} \\
    \wh{\sf b} &=& \sin\beta\;\wh{\sf z} \;-\; \cos\beta\,\wh{\varphi}, \label{eq:b_ab} 
\end{eqnarray}
where we introduced the spherical angles $(\alpha,\beta)$: 
\begin{equation}
    \left. \begin{array}{rcl}
    A &\equiv& \cos\alpha\;\cos\beta \\
    B &\equiv& \sin\alpha\;\cos\beta \\
    C &\equiv& \sin\beta
    \end{array} \right\}.
\end{equation} 
As a special case, the planar (2D) elastica curve, which corresponds to the torsionless case $\nu = 0$ (i.e., $\beta = 0$), is presented in App.~\ref{sec:planar}.

\subsection{Periodic vertical solution $z(\xi;m)$}

We begin by solving the vertical equation (\ref{eq:z_prime}). Using Eq.~(\ref{eq:A_def}), we integrate Eq.~(\ref{eq:z_prime}) to obtain
\begin{eqnarray}
    z(\xi) &=& {\cal R}^{2}k_{0}\,\left(1 + \frac{\gamma^{2}}{2}\right)\, \xi \;-\; \frac{{\cal R}^{2}k_{0}}{\sqrt{q_{0}}} \int_{0}^{\xi/\sqrt{q_{0}}}\;m\;
    {\rm sn}^{2}(u|m)\;du,
    \label{eq:zs_def}
\end{eqnarray}
where we used the initial condition $z(0) = 0$, and we initially assume that $(m,q_{0})$ are both positive. 

We now introduce the periodic Jacobi zeta function \cite{AS} ${\cal Z}(u|m) \equiv {\sf Z}[{\rm am}(u|m)\,|\,m]$, with the Jacobi amplitude function ${\rm am}(u|m) \equiv {\rm arcsin}[{\rm sn}(u|m)]$, and we use the identity \cite{AS}
\[ m\,\int_{0}^{v}{\rm sn}^{2}(u|m)\,du \;=\; \left(1 - \frac{{\sf E}(m)}{{\sf K}(m)}\right)\,v \;-\; {\cal Z}(v|m), \]
where ${\cal Z}(v + 2\,{\sf K}(m)|m) = {\cal Z}(v|m)$. Here, the complete elliptic integrals ${\sf K}(m)$ and ${\sf E}(m)$ of the first and second kinds are defined in Eqs.~(\ref{eq:K_m}) and (\ref{eq:E_m}), respectively.

\begin{figure}
\epsfysize=1.7in
\epsfbox{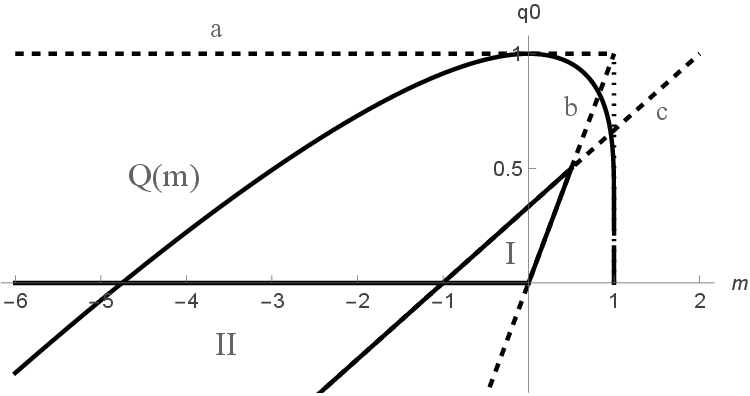}
\caption{Plot of $Q(m)$ (obtained from the periodicity constraint $\Delta z = 0$) drawn on the plane $(m,q_{0})$ in the range $-6 \leq m \leq 1$, with the curves a, b, and c shown from Fig.~\ref{fig:gammanu_2}. We see that the curve $Q(m)$ completely misses region I and enters region II for $q_{0} < 0$ and $m < m_{0}^{-} \simeq -4.7510$.}
\label{fig:Q0_m}
\end{figure}

Starting with the case $q_{0} > 0$, Eq.~(\ref{eq:zs_def}) yields the preliminary solution for the vertical position
\begin{eqnarray}
    z(\xi) &=& {\cal R}^{2}k_{0}\,\left[ \left(1 + \frac{\gamma^{2}}{2}\right) \;+\; \frac{1}{q_{0}}\left( \frac{{\sf E}(m)}{{\sf K}(m)} \;-\; 1\right)\right]\,\xi \;+\; \frac{{\cal R}^{2}k_{0}}{\sqrt{q_{0}}}\;
    {\cal Z}\left(\left.\frac{\xi}{\sqrt{q_{0}}} \right|\;m\right),
    \label{eq:z_prelim}
\end{eqnarray}
which is represented as the sum of a term that increases linearly with $\xi$ and a periodic term ${\cal Z}(\xi/\sqrt{q_{0}}|\;m)$ with the period $\Xi_{z} = 4\,\sqrt{q_{0}}\,{\sf K}(m)$. Hence, if the coefficient of the linear term does not vanish, Eq.~(\ref{eq:z_prelim}) represents an open helical spatial curve.

\subsubsection{Closed elastica knot}

Since a closed knot must be confined in space, we want the vertical solution to be periodic for all $(m,q_{0})$:
\begin{equation}
    \Delta z \;=\; z(\xi + \Xi_{z}) - z(\xi) \;\equiv\; 0,
    \label{eq:delta_z}
\end{equation}
which requires that
\begin{eqnarray} 
0 &=& 1 + \frac{1}{2}\,\gamma^{2}(m,q_{0}) + \frac{1}{q_{0}}\left( \frac{{\sf E}(m)}{{\sf K}(m)} - 1\right) \;=\; \frac{(1 + m)}{2q_{0}} - \frac{1}{2} + \frac{1}{q_{0}}\left( \frac{{\sf E}(m)}{{\sf K}(m)} - 1\right),
 \end{eqnarray}
where we substituted Eq.~(\ref{eq:lambda_mq0}). Hence, the vertical periodicity (\ref{eq:delta_z}) imposes the following constraint on $q_{0}$:
\begin{eqnarray}
q_{0} \;=\; Q(m) &\equiv& 2\,\frac{{\sf E}(m)}{{\sf K}(m)} \;-\; (1 - m) \;=\; 2 \left(\frac{{\sf E}(m)}{{\sf K}(m)} - 1\right) \;+\; (1 + m),
\label{eq:q_LS}
\end{eqnarray}
where $Q(0) = 1$, while $Q(m)$ vanishes at  $m = 1$ and $m = m_{0}^{-} \simeq -4.751$. We note that Eq.~(\ref{eq:q_LS}), which can be rewritten as $1 + q_{0} - m - 2\,{\sf E}(m)/{\sf K}(m) = 0$, is identical to the periodicity condition obtained by Langer and Singer \cite{Langer_Singer_1984}, which is now extended to a larger parameter range $(m,q_{0})$, since $m$ and/or $q_{0}$ are not constrained to be positive.

\begin{figure}
\epsfysize=1.5in
\epsfbox{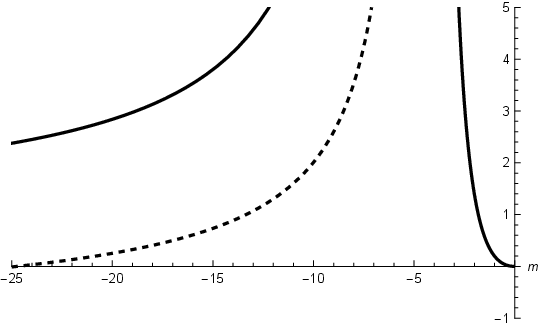}
\caption{Plots of $\nu^{2}(m)$ (solid) and $\gamma^{2}(m)$ (dashed) versus $m$ in the range $-25 \leq m \leq 0$. We note that both functions are simultaneously positive for $m < m_{0}^{-} \simeq -4.751$, while they diverge at $m = m_{0}^{-} \simeq -4.751$ (shown as a dotted vertical line). Moreover, $\gamma^{2}(m)$ (dashed) becomes negative for $m < m_{1}^{-} \simeq -24.74$, while $\nu^{2}(m)$ remains positive.}
\label{fig:Gamma_Nu}
\end{figure}

\subsubsection{Extended elastica-knot parameter space}

Figure \ref{fig:Q0_m} shows the function $Q(m)$ in the $(m,q_{0})$-plane in the range $-6 \leq m \leq 1$, with the dashed lines (a,b,c) defined in Fig.~\ref{fig:gammanu_2}. On this curve, the torsion function (\ref{eq:nu2_mq0}) becomes a function of $m$ alone:
\begin{equation}
    \nu^{2}[m,Q(m)] \;=\; \frac{[1 - Q(m)]\,[Q(m) - m]}{Q^{2}(m)},
    \label{eq:nu2_Q0}
\end{equation}
which is positive for $m < Q(m) < 1$. We note that $Q(m)$ reaches the boundaries $q_{0} = 1$ (curve a) and $q_{0} = m$ (curve b) at $m = 0$ and $m_{0}^{+} \simeq 0.8261$ (related to $m_{0}^{-}$ as $m_{0}^{+} = -m_{0}^{-}/(1 - m_{0}^{-})$), respectively. If we also insert $Q(m)$ into Eq.~(\ref{eq:q0_m}), we obtain the wave-speed function
\begin{equation}
    \gamma^{2}(m) \;=\; \frac{(1 + m)}{Q(m)} \;-\; 3,
    \label{eq:gamma2_Q0}
\end{equation}
which is positive for $m_{1}^{+} \simeq 0.96115 < m < 1$ (outside region I) and for $m_{1}^{-} \simeq -24.7397 < m < m_{0}^{-} \simeq -4.751 < 0$ (inside region II in Fig.~\ref{fig:Q0_m}). Figures \ref{fig:Q0_m} and \ref{fig:Gamma_Nu}, therefore, show that the traveling-wave NLSE solution cannot be found in the classical knot parameter range $0 \leq m \leq q_{0} \leq 1$, and can only be found in the extended parameter space (region II in Fig.~\ref{fig:Q0_m}), where $m \leq q_{0} = Q(m) \leq m_{0}^{-} < 0$. More importantly, we clearly see from Fig.~\ref{fig:Q0_m} and Eq.~(\ref{eq:m_extended}) that the classically forbidden condition \cite{LS_1984}
\begin{equation}
2 \;+\; \gamma^{2}(m) \;-\; \nu^{2}(m) \;=\; m/Q^{2}(m) \;<\; 0
\label{eq:m_forbidden}
\end{equation} 
can, in fact, be satisfied.

\begin{figure}
\epsfysize=1.7in
\epsfbox{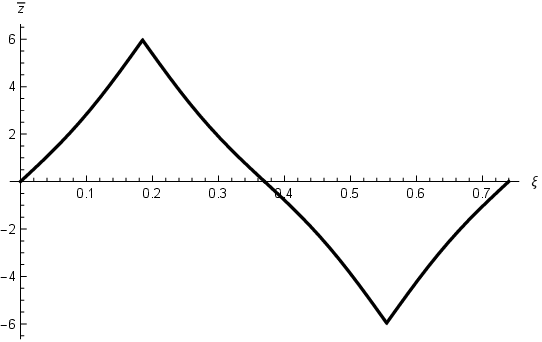}
\caption{Plot of the normalized vertical solution (\ref{eq:z_sol}) versus $\xi$ for $m = -6$, over one period (\ref{eq:S_z}).}
\label{fig:zeta}
\end{figure}

The special case where $\gamma^{2}(m)$ vanishes corresponds to the condition $Q(m_{1}^{\pm}) = (1 + m_{1}^{\pm})/3$, or
\[ {\sf E}(m_{1}^{\pm})/{\sf K}(m_{1}^{\pm}) \;=\; (2 - m_{1}^{\pm})/3, \]
which yields $m_{1}^{-} \simeq -24.74 \equiv -\,m_{1}^{+}/(1 = m_{1}^{+})$. For these values, the speed of the traveling wave $c(m_{1}^{\pm}) = k_{0}D\,\gamma(m_{1}^{\pm}) = 0$ vanishes, and $\Omega^{2}(m_{1}^{\pm}) = \nu^{2}(m_{1}^{\pm})/[1 + \nu^{2}(m_{1}^{\pm})] < 1$.

\subsubsection{Periodic vertical solution}

The periodic vertical solution (\ref{eq:z_prelim}) is expressed in dimensionless form $\ov{z}(\xi) \equiv k_{0}\,z(\xi)$ as
\begin{equation}
    \ov{z}(\xi;m) = \frac{{\cal R}^{2}(m)\,k_{0}^{2}}{\sqrt{Q(m)}}\;
    {\cal Z}\left(\left.\frac{\xi}{\sqrt{Q(m)}}\right|\;m\right),
    \label{eq:z_sol}
\end{equation}
where $Q(m)$ is given by Eq.~(\ref{eq:q_LS}). The requirement of periodicity and the positivity of the functions (\ref{eq:nu2_Q0}) and (\ref{eq:gamma2_Q0}) implies that the Jacobi parameter must be in the range $m < m_{0}^{-} \simeq -4.751$ (region II in Fig.~\ref{fig:Q0_m}), where $Q(m) < 0$. Here, Fig.~\ref{fig:zeta} shows that the function $i\,{\cal Z}(i\,y|m)$ is real and has a finite period $4\,\sqrt{n'}\,{\sf K}(n^{\prime})$ for $m < 0$, which implies that the periodic vertical solution (\ref{eq:z_sol}) has a period 
\begin{equation} 
    \Xi_{z}(m) \equiv 4\,\sqrt{|Q(m)|n'(m)}\;{\sf K}[n'(m)],
    \label{eq:S_z}
\end{equation}
which vanishes at $m = m_{0}^{-}$.

\subsection{Periodic radial solution $\rho(\xi;m)$}

\begin{figure}
\epsfysize=1.7in
\epsfbox{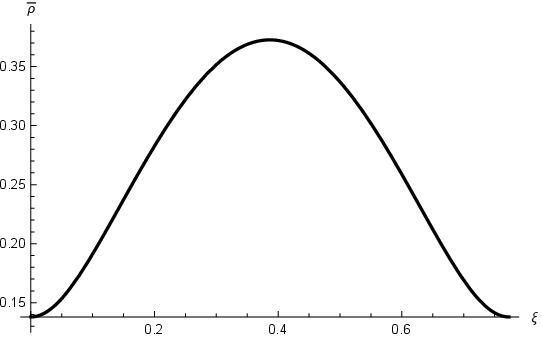}
\caption{Plot of the normalized radial solution (\ref{eq:rho_sol}) versus $\xi$ for $m = -6$, over one radial period (\ref{eq:S_rho}).}
\label{fig:rho_xi}
\end{figure}

Next, we derive the radial solution from Eq.~(\ref{eq:rho_prime}). Using Eq.~(\ref{eq:B_def}), the radial derivative (\ref{eq:rho_prime}) becomes
\begin{eqnarray}
    \rho^{\prime}(s) &=& \frac{{\cal R}^{2}\,\kappa(s)\,\kappa^{\prime}(s)}{\sqrt{\kappa^{2}(s) \;-\; k_{0}^{2}\Omega^{2}}} \;\equiv\; \frac{d}{ds}\left({\cal R}^{2}\;\sqrt{\kappa^{2}(s) \;-\; k_{0}^{2}\,\Omega^{2}}\right),
\end{eqnarray}
which can be integrated exactly to yield the intrinsically periodic solution
\begin{equation}
    \rho(s) \;=\; {\cal R}^{2}\;\sqrt{\kappa^{2}(s) \;-\; k_{0}^{2} \,\Omega^{2}},
\end{equation}
where, since $1 - \Omega^{2} > 0$, the initial radial position is $\rho_{0}(m) = {\cal R}^{2}(m)\,k_{0}\,\sqrt{1 - \Omega^{2}(m)}$. Hence, the normalized radial position $\ov{\rho}(\xi) \equiv k_{0}\,\rho(\xi)$ is solved as
\begin{equation}
    \ov{\rho}(\xi;m) = {\cal R}^{2}\,k_{0}^{2}\,\sqrt{(1 - \Omega^{2}) - \frac{m}{Q(m)}\;{\rm sn}^{2}\left(\left.\frac{\xi}{\sqrt{Q(m)}}\right|\;m\right)},
    \label{eq:rho_sol}
\end{equation}
which has a maximum ${\cal R}^{2}\,k_{0}^{2}\,\sqrt{1 - \Omega^{2}}$ at $\xi = 0$. On the other hand, in the extended range $m < m_{0}^{-}$, with $q_{0} = Q(m) < 0$, we use Eq.~(\ref{eq:sn_y}) so that the normalized radial solution (\ref{eq:rho_sol}) becomes
\begin{equation}
    \ov{\rho}(\xi;m) = {\cal R}^{2}\,k_{0}^{2}\,\sqrt{(1 - \Omega^{2}) + \frac{n\,{\rm sd}^{2}(\xi/\sqrt{|Q(m)|\,n'}|\;n')}{|Q(m)|}}.
    \label{eq:radial_ext}
\end{equation}
This radial solution, shown in Fig.~\ref{fig:rho_xi} for $m = -6$, is periodic with a period that is half of the period (\ref{eq:S_z}) for the vertical solution (\ref{eq:z_sol}):
\begin{equation} 
\Xi_{\rho} \;\equiv\; \Xi_{z}/2 \;=\; 2\,{\sf K}[n'(m)]\,\sqrt{|Q(m)|n'(m)},
\label{eq:S_rho}
\end{equation}
and the extended radial solution (\ref{eq:radial_ext}) has a maximum ${\cal R}^{2}\,k_{0}^{2}\,\sqrt{(1 - \Omega^{2}) + 1/|Q|}$ at the half-period ${\sf K}(n')\,\sqrt{|Q|n'}$.

\subsection{\label{sec:angle}Aperiodic azimuthal angular solution $\varphi(s;m)$}

When the radial solution (\ref{eq:rho_sol}) is inserted into the angular equation (\ref{eq:varphi_prime}), we obtain
\begin{eqnarray}
    \varphi^{\prime}(s) &=& \frac{1}{2}\,{\cal R}^{2}k_{0}^{2}\,\tau_{0} \left( \frac{\kappa^{2}(s) \;+\; \frac{1}{2}\, \gamma^{2}\,k_{0}^{2}}{\kappa^{2}(s) \;-\; \Omega^{2}\,k_{0}^{2}}\right) \;\equiv\; \frac{k_{0}\Omega}{2} \left[1 \;+\; \frac{k_{0}^{2}\,(\Omega^{2} + \frac{1}{2}\,\gamma^{2})}{\kappa^{2}(s) \;-\; \Omega^{2}\,k_{0}^{2}} \right] \nonumber \\
    &=& \frac{k_{0}\Omega}{2} \left[ 1 \;+\; 
    \frac{(\Omega^{2} + \frac{1}{2}\,\gamma^{2})}{(1 - \Omega^{2}) - (m/q_{0})\;{\rm sn}^{2}(\xi/\sqrt{q_{0}}|m)} \right], \nonumber 
\end{eqnarray}
where we used the definition (\ref{eq:Omega_def}) for $\Omega$, and we explicitly assume that $q_{0} \equiv Q(m)$ in what follows. This equation can be solved as
\begin{equation}
    \varphi(\xi;m) = \Omega \,\xi + \int_{0}^{\xi}\,
    \frac{\Omega \,(\Omega^{2} + \frac{1}{2}\,\gamma^{2})\;dv}{(1 - \Omega^{2}) - (m/q_{0})\;{\rm sn}^{2}(v/\sqrt{q_{0}}|m)},
    \label{eq:varphi_sol_1}
\end{equation}
Instead of solving the integral in terms of the incomplete elliptic integral of the third kind (see Case (iii) in Sec. 3.7 in Lawden's book \cite{Lawden}), we now proceed with the simpler solution based on the Weierstrass elliptic functions.

First, we use Eq.~(\ref{eq:sn_P}) to introduce the new parameter function $u(m,q_{0})$:
\begin{equation}
   1 - \Omega^{2} \equiv (m/q_{0})\;{\rm sn}^{2}(u/\sqrt{q_{0}}|m)
   \equiv \wp(u + \omega_{3}) - {\sf e}_{3},
    \label{eq:snP_u}
\end{equation}
where
\begin{eqnarray}
    u(m,q_{0}) &\equiv& \sqrt{q_{0}}\,{\rm sn}^{-1}\left(\left.
    \sqrt{\frac{q_{0}}{m}\,(1 - \Omega^{2})}\,\right|\,m\right) \;=\; \sqrt{q_{0}}\,{\rm F}(\phi'|m)
\end{eqnarray}
is expressed in terms of the incomplete elliptic integral of the first kind $F(\phi'|m)$, defined in Eq.~(\ref{eq:v_Fphi}), where 
\[ \sin\phi' \;\equiv\; \sqrt{q_{0}\,(1 - \Omega^{2})/m} < 1 \]
for $(m,q_{0})$ in region II. We note that, in the range where both $(m,q_{0})$ are negative, the function $u(m,q_{0})$ is imaginary.

Second, by using the relation $\wp(u + \omega_{3}) = (1 - \Omega^{2}) + {\sf e}_{3}$, we obtain
\begin{eqnarray}
    [\wp^{\prime}(u + \omega_{3})]^{2} &=& 4\,[\wp(u + \omega_{3})]^{3} \;-\; {\sf g}_{2}\,\wp(u + \omega_{3}) \;-\; {\sf g}_{3} \nonumber \\
      &=& 4\,(1 - \Omega^{2})^{3} + 12\,{\sf e}_{3}\,(1 - \Omega^{2})^{2} \;+\; 12\,\left({\sf e}_{3}^{2} - {\sf g}_{2}\right)\,(1 - \Omega^{2}),
\end{eqnarray}
where we used the fact that ${\sf e}_{3}$ is a root of the cubic polynomial $4\,z^{3} - {\sf g}_{2}\,z - {\sf g}_{3}$. After several manipulations, which include using the relations $12\,{\sf e}_{3}^{2} \;-\; {\sf g}_{2} = 4\,m/q_{0}^{2} = 8 + 4\,(\gamma^{2} - \nu^{2})$ and
\[ \nu^{2} \;=\; \frac{\Omega^{2}}{1 - \Omega^{2}} \left(1 + \frac{1}{2}\,\gamma^{2}\right)^{2}, \]
we finally obtain $[\wp^{\prime}(u + \omega_{3})]^{2} = -\,4\,
\Omega^{2} (\Omega^{2} + \gamma^{2}/2)^{2}$, which yields
\begin{equation}
    \left(\Omega^{2} \;+\; \frac{1}{2}\,\gamma^{2}\right) \;=\;
    \frac{-i}{2\Omega}\;\wp^{\prime}(u + \omega_{3}).
\end{equation}
Hence, Eq.~(\ref{eq:varphi_sol_1}) becomes
\begin{equation}
    \varphi(\xi;m) \;=\; \Omega\,\xi \;+\; \frac{i}{2} \int_{0}^{\xi} 
    \frac{\wp^{\prime}(u + \omega_{3})\;dv}{\wp(v + \omega_{3}) \;-\;
    \wp(u + \omega_{3})}.
    \label{eq:varphi_sol_2}
\end{equation}
Next, we use the identity \cite{NIST_Chap23}
\begin{equation} 
\frac{\wp^{\prime}(y)}{\wp(x) - \wp(y)} \;=\; 2\,\zeta(y) \;+\; 
\frac{d}{dx}\ln\left(\frac{\sigma(x-y)}{\sigma(x+y)}\right),
\label{eq:wp_zeta_sigma}
\end{equation}
where the Weierstrass elliptic functions $\sigma(z)$ and $\zeta(z)$ are defined in terms of the Weierstrass elliptic function $\wp(z)$ as
$\wp(z) \equiv -\,\zeta^{\prime}(z)$ and $\zeta(z) \equiv \sigma^{\prime}(z)/\sigma(z)$. The integral in Eq.~(\ref{eq:varphi_sol_2}) can now be solved exactly, and we finally obtain
\begin{eqnarray}
    \varphi(\xi;m) &=& \left(\Omega \;+\; i\;\left[\zeta(u + \omega_{3}) - \zeta(\omega_{3})\right]\right)\,\xi \;+\; \frac{i}{2} \ln\left(\frac{\sigma(u - \xi)}{\sigma(u + \xi)}\right),
    \label{eq:varphi_sol_3}
\end{eqnarray}
where we used the quasi-periodic properties of the Weierstrass sigma function \cite{Lawden}
\begin{equation}
    \sigma(z \pm 2\,\omega_{3}) \;\equiv\; -\,\sigma(z)\;
    \exp\left[ \pm\,2\,\eta_{3}\,(z \pm \omega_{3})\right],
    \label{eq:sigma_period}
\end{equation}
where $\eta_{3} \equiv \zeta(\omega_{3})$. We note, here, that since 
$\xi$ is real and $u$ is imaginary, the ratio 
\begin{equation} 
\frac{\sigma(u-\xi)}{\sigma(u + \xi)} = -\,\frac{\sigma(\xi-u)}{\sigma(\xi + u)} \;\equiv\; \exp\left[ i\,\pi \;-\; 2\,i\,\chi(\xi;u)\right]
\end{equation}
has a unit magnitude, where $\chi(\xi;u)$ is defined as the argument of $\sigma(\xi + u)$. Hence, the azimuthal angle (\ref{eq:varphi_sol_3}) can also be written as
\begin{eqnarray}
    \varphi(\xi;m) &=& \left(\Omega \;+\; i\;\left[\zeta(u + \omega_{3}) - \zeta(\omega_{3})\right]\right)\,\xi \;+\; \chi(\xi;u) \;-\; \frac{\pi}{2},
    \label{eq:varphi_sol_4}
\end{eqnarray}
which is convenient for numerical purposes.

\begin{figure}
\epsfysize=2in
\epsfbox{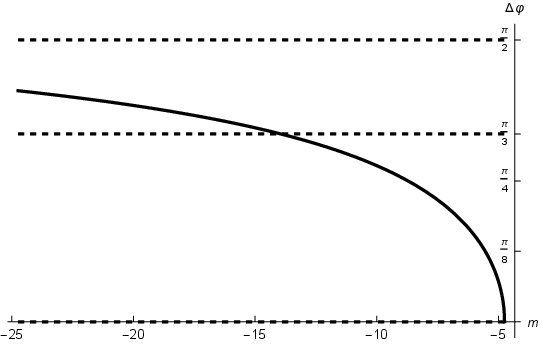}
\caption{Plot of $\Delta\varphi(m) = \varphi(\xi + 2\omega_{1}) - \varphi(\xi)$ as a function of $m$ in the range $m_{1}^{-} < m < m_{0}^{-}$. For all values in that range, we find $\Delta\varphi(m) < \pi/2$. We selected $\Delta\varphi(m_{2}) = \pi/3$ (represented as a dashed horizontal line) at $m_{2} \simeq -13.9483$ to generate the closed elastica knot represented in Figs.~\ref{fig:XYZ_3D}-\ref{fig:XYZ_knot}.}
\label{fig:Delta_phi}
\end{figure}

Finally, for $q_{0} = Q(m) < 0$ (region II in Fig.~\ref{fig:Q0_m}), where $u$ is imaginary, the elliptic functions $i\;\left[\zeta(u + \omega_{3}) - \zeta(\omega_{3})\right]$ and $i\,\ln[\sigma(u - \xi)/\sigma(u + \xi)]$ are both real. Moreover, the latter function is periodic, with period $2\sqrt{q_{0}}\,{\sf K}(m) \equiv 2\,\omega_{1}$. Hence, after a single period, the azimuthal angular shift $\Delta\varphi(m) \equiv \varphi(\xi + 2\omega_{1}) - \varphi(\xi)$ is expressed as
\begin{eqnarray}
   \Delta\varphi(m) &=& 2i\, \left[\zeta(u + \omega_{3}) - \zeta(\omega_{3})\right]\,\omega_{1} \;+\; 2\,\Omega\,\omega_{1} \;-\; 2i\,\zeta(\omega_{1})\,u \nonumber \\
   &=& 2i\, \left[\zeta(u + \omega_{3})\,\omega_{1} - \zeta(\omega_{1})\,(u + \omega_{3})\right] \;+\; 2\,\Omega\,\omega_{1} \;-\; \pi,
   \label{eq:Delta_phi}
\end{eqnarray}
where we used 
\[ \chi(\xi+ 2\omega_{1};u) \;=\; \chi(\xi;u) \;-\; 2\,i\,\zeta(\omega_{1})\,u, \]
and the identity $\zeta(\omega_{3})\,\omega_{1} - \zeta(\omega_{1})\,\omega_{3} \equiv -i\,\pi/2$. We note that the elastica knot is closed only if $\Delta\varphi$ is equal to a rational fraction of $\pi$. Figure \ref{fig:Delta_phi} shows that $0 \leq \Delta\varphi(m) < \pi/2$ for $m \leq m_{0}^{-} \simeq -4.751$ and all rational fractions of $\pi$ smaller than $\pi/2$ are reached at specific values of $m$. 

Lastly, using the definition $\wp(z) \equiv -\,\zeta^{\prime}(z)$, the angular shift (\ref{eq:Delta_phi}) can be compactly expressed as
\begin{equation}
    \Delta\varphi(m) \;\equiv\; 2\,\Omega\,\omega_{1} \;-\; 2\,\int_{0}^{iu}\left[\omega_{1}\,\wp(\omega_{3} - i\,t) \;+\; \zeta(\omega_{1})\right]\,dt,
    \label{eq:delta_phi_W}
\end{equation}
which is a form also used in the elliptic solution of the spherical pendulum \cite{Brizard_2009}.

\subsection{Closed 3D elastica knot}

\begin{figure}
\epsfysize=4.5in
\epsfbox{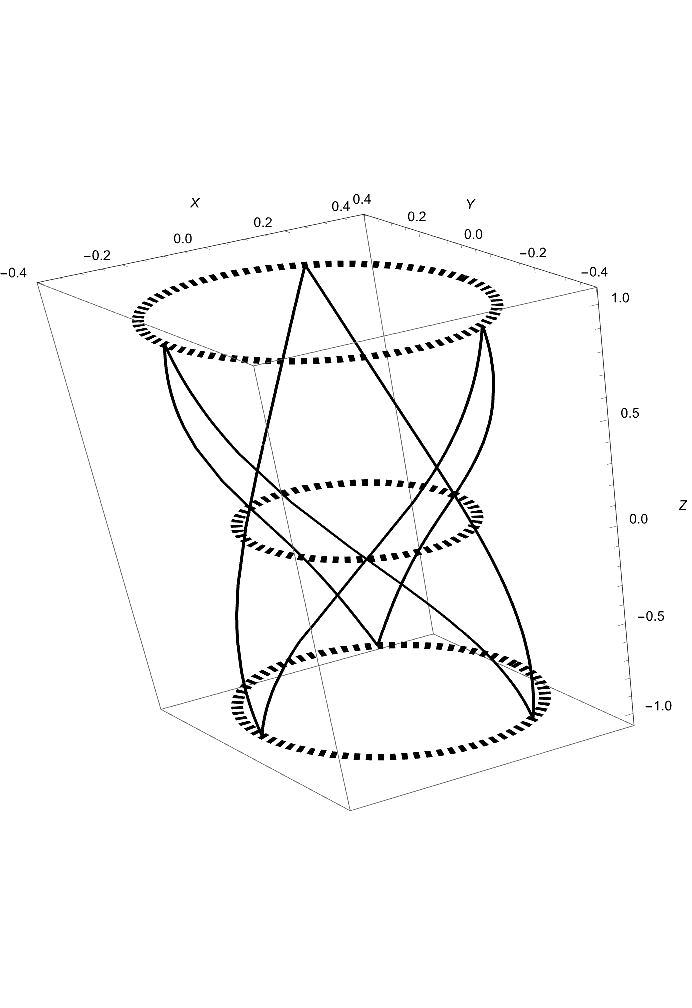}
\caption{Parametric 3D plot of the elastica knot $(\ov{x},\ov{y},\ov{z})$ in the range $0 \leq \xi \leq 12\,\omega_{1}(m_{2})$ for $m_{2} \simeq -13.9483$, corresponding to the case $\Delta\varphi(m_{2}) = \pi/3$. Dashed circles at top and bottom are drawn at maximum radius $\ov{\rho}(\omega_{1})$ and top/bottom vertical positions $\pm\ov{z}(\omega_{1})$, and at center drawn at minimumn radius $\ov{\rho}(0)$ and $\ov{z}(0) = 0$.}
\label{fig:XYZ_3D}
\end{figure}

\begin{figure}
\epsfysize=2.4in
\epsfbox{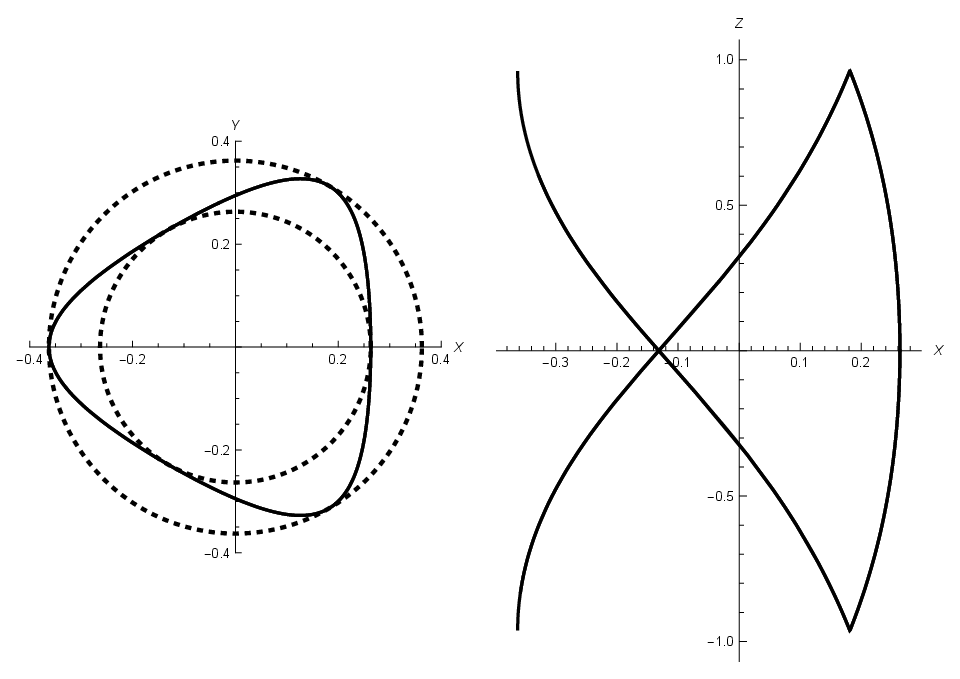}
\caption{Top view (left) and side view (right) of the elastica knot (with $\Delta\varphi(m_{2}) = \pi/3$) shown in Fig.~\ref{fig:XYZ_3D}. The minimum and maximum normalized radii are shown as dashed circles in the left plot.}
\label{fig:XYZ_knot}
\end{figure}

Using the plot of $\Delta\varphi(m)$ shown in Fig.~\ref{fig:Delta_phi}, we select the closed 3D elastica knot corresponding to $\Delta\varphi = \pi/3$ (represented as a dashed horizontal line in Fig.~\ref{fig:Delta_phi}), which occurs at the extended parameter $m_{2} \simeq -13.9483$. Figure \ref{fig:XYZ_3D} shows the torus elastica knot, where the normalized Cartesian coordinates $(\ov{x},\ov{y},\ov{z})$ are plotted for 6 periods, i.e., $6\times 2\,\omega_{1}(m_{2}) = 12\,\omega_{1}(m_{2})$ corresponds to $6\times\pi/3 = 2\pi$. Dashed circles at top and bottom are drawn at maximum radius $\ov{\rho}(\omega_{1})$ and top/bottom vertical positions $\pm\ov{z}(\omega_{1})$, and at center drawn at minimumn radius $\ov{\rho}(0)$ and $\ov{z}(0) = 0$. Figure \ref{fig:XYZ_knot}, on the other hand, shows the top view (left) of Fig.~\ref{fig:XYZ_3D}: $\ov{y}(\xi) \equiv \ov{\rho}(\xi)\,\sin\varphi(\xi)$ versus $\ov{x}(\xi) \equiv \ov{\rho}(\xi)\,\cos\varphi(\xi)$ and the side view (right) of Fig.~\ref{fig:XYZ_3D}: $\ov{z}(\xi)$ versus $\ov{x}(\xi)$. 

We, therefore, conclude that a closed elastica knot, with $\Delta\varphi(m_{c}) \equiv p_{1}\pi/p_{2}$ defined in terms of two integers $p_{2} > 2p_{1}$, occurs at a specific value $m_{c}(p_{1},p_{2}) < m_{0}^{-}$, with $q_{0c} \equiv Q(m_{c}) < 0$.

\section{\label{sec:sec_7}NLSE on a 3D elastica knot}

In this Section, we will use our elliptic solutions (\ref{eq:elastica_J}) and (\ref{eq:W_kappa}) of the curvature equation (\ref{eq:kappa_elastica}) to obtain solutions of the NLSE (\ref{eq:NLSE}). First, we will consider a separable solution that is periodic in $s$ and, then, we will seek traveling-wave solutions.

\subsection{\label{sec:Lame}Periodic Lam\'{e} solution}

Now that we have derived the solution (\ref{eq:elastica_J}) for the squared curvature $\kappa^{2}(s)$, it is instructive to return to the NLSE (\ref{eq:NLSE}), which can be expressed as a linear Schr\"{o}dinger equation for a particle of mass $M$ traveling in a time-independent potential $V(s) \equiv -\,\hbar^{2}\kappa^{2}(s)/(4M)$:
\begin{equation}
    \frac{-i}{D}\pd{\psi}{t} = \frac{\partial^{2}\psi}{\partial s^{2}} + \frac{k_{0}^{2}}{2}\left[1 - \frac{m}{q_{0}}\,{\rm sn}^{2}\left(\left. 
    \frac{k_{0}s}{2\sqrt{q_{0}}}\right|m\right)\right]\psi.
    \label{eq:psi_kappa}
\end{equation}
Next, we consider the transformation to the Lam\'{e} equation \cite{Ince_1940,Erdelyi_1941,NIST_Chap29}
\begin{equation}
    \frac{d^{2}y(u)}{du^{2}} \;=\; \left[\ell (\ell+1)\,m\,{\rm sn}^{2}(u|m) - h\right]\,y(u),
    \label{eq:Lame_eq}
\end{equation}
where the index $\ell \geq -1/2$ may be an integer and $h$ is the eigenvalue associated with a periodic solution $y(u)$, with period $2\,{\sf K}(m)$ or $4\,{\sf K}(m)$. 

We begin with the case where both knot parameters $(m,q_{0})$ are positive. When we substitute 
\begin{equation}
    \psi(s,t) \;\equiv\; \Psi(u)\,\exp\left(iD\,\epsilon^{2}k_{0}^{2}t/2\right) 
\end{equation}
into Eq.~(\ref{eq:psi_kappa}), where $u \equiv k_{0}s/(2\sqrt{q_{0}})$ and $\epsilon$ is a real parameter, we obtain the second-order ordinary differential equation
\begin{eqnarray}
    \Psi^{\prime\prime}(u) &=& 2\,m\;{\rm sn}^{2}(u|m)\;\Psi(u) \;-\; 2\,q_{0}\left(1 - \epsilon^{2}\right)\,\Psi(u).
    \label{eq:Lame_NLSE}
\end{eqnarray}
By comparing this equation with the Lam\'{e} equation (\ref{eq:Lame_eq}), we easily solve the index equation $2 = \ell\,(\ell + 1)$ with the integer $\ell = 1$, which implies that the solution $\Psi(u)$ is a first-order polynomial in a Jacobi elliptic function \cite{Ince_1940,Erdelyi_1941}. Since ${\rm dn}(u|m)$ satisfies the second-order differential equation 
\[ \frac{d^{2}{\rm dn}(u|m)}{du^{2}} \;=\; \left[2\,m\,{\rm sn}^{2}(u|m) - m\right]\,
{\rm dn}(u|m), \]
we easily find that Eq.~(\ref{eq:Lame_NLSE}) has the periodic solution 
\begin{equation} 
    \Psi(u) \;=\; A\,{\rm dn}(u|m),
\end{equation}
with an arbitrary amplitude $A$ and a period $2\,{\sf K}(m)$. Here, the Jacobi parameter $m$ is defined by the eigenvalue equation $m \equiv 2\,q_{0}\,(1 \;-\; \epsilon^{2}) \leq 1$, which yields
\begin{equation}
    \epsilon^{2}(m,q_{0}) \;=\; 1 \;-\; m/(2q_{0}) \geq 0,
    \label{eq:epsilon2}
\end{equation}
and, thus, Eq.~(\ref{eq:epsilon2}) is satisfied since $q_{0} \geq m > m/2$. Hence, the periodic solution of the NLSE (\ref{eq:NLSE}) is expressed as
\begin{equation}
    \psi(s,t) \equiv k_{0}\,{\rm dn}\left(\left.\frac{k_{0}s}{2\sqrt{q_{0}}}\right|m\right)\,\exp\left[\frac{iD}{2}\,\epsilon^{2}(m,q_{0})\;k_{0}^{2}\,t\right],
    \label{eq:psi_dn}
\end{equation}
where the amplitude $k_{0}$ is chosen to match the curvature, and the period $S(m,q_{0})$ is defined as $k_{0}S(m,q_{0})/2 \equiv 2\,\sqrt{q_{0}}\,{\sf K}(m)$. When $\epsilon = 1/\sqrt{2}$, for example, we obtain the torsionless case $q_{0} = m$ which, for $m = 1$, yields the optical-soliton solution (\ref{eq:optical_soliton}):
\begin{equation}
    \psi(s,t) \;=\; 2k\;{\rm sech}(ks)\;\exp\left(iD\,k^{2}t\right),
\end{equation}
whose amplitude $2k = k_{0}$ is chosen to match the curvature $\kappa(s) = k_{0}\;{\rm sech}(k_{0}s/2)$.

Finally, we note that the periodic NLSE solution (\ref{eq:psi_dn}) is still valid for negative values of $(m,q_{0})$. Indeed, for $(m<0,q_{0}<0)$, we use the identity (\ref{eq:dn_y}), to find
\begin{equation}
    {\rm dn}\left(\left.\frac{k_{0}s}{2\sqrt{q_{0}}}\right|m\right) \;=\; {\rm cd}\left(\left.\frac{k_{0}s}{2\sqrt{|q_{0}|\,n'}}\right|n'\right),
\end{equation}
where the new Jacobi parameter $n' = 1/(1 - m) < 1$ falls in the classical range, and the periodic solution has a period $S(m,q_{0})$ defined as $k_{0}S(m,q_{0})/2 \equiv 4\,\sqrt{|q_{0}|\,n'}\,{\sf K}(n')$.

\subsection{\label{sec:TW_NLSE}Traveling-wave NLSE solutions} 

We now come back to the traveling-wave NLSE solution (\ref{eq:ansatz}). Once again, according to Fig.~\ref{fig:Q0_m}, this solution can only be found in region II (see Fig.~\ref{fig:Q0_m}) of the extended parameter space $(m,q_{0})$, where $m_{1}^{-} \simeq -24.74 < m < m_{0}^{-} \simeq -4.75 < 0$ and $q_{0} = Q(m)$ are both negative, which corresponds to both functions (\ref{eq:nu2_Q0})-(\ref{eq:gamma2_Q0}) being positive (see Fig.~\ref{fig:Gamma_Nu}).

\subsubsection{Jacobi elliptic solution} 

The Jacobi elliptic traveling-wave solution of the NLSE (\ref{eq:NLSE}) on an elastica knot proceeds as follows. First, we write the ansatz (\ref{eq:ansatz}) in terms of the elastica solution (\ref{eq:elastica_J}):
\begin{equation}
    \Psi(\xi_{t}) \;\equiv\; k_{0}\,\sqrt{1 \;-\; \frac{m}{q_{0}}\;{\rm sn}^{2}\left(\left.\frac{\xi_{t}}{\sqrt{q_{0}}}\right|m\right)}\;e^{i\,\theta(\xi_{t})},
    \label{eq:ansatz_sol}
\end{equation}
where we defined $\xi_{t} \equiv k_{0}s_{t}/2$ and the phase $\theta(\xi_{t})$ is defined as [compare with Eq.~(\ref{eq:varphi_sol_1})]
\begin{equation}
    \theta(\xi_{t}) \;\equiv\; \gamma\,\xi_{t} \;+\; \int_{0}^{\xi_{t}}\frac{\nu\;dv}{1 \;-\; (m/q_{0})\;{\rm sn}^{2}(v/\sqrt{q_{0}}|m)},
    \label{eq:Phi_def}
\end{equation}
where the integral can be solved \cite{Barros_2018} in terms of the incomplete elliptic integral of the third kind. We note that this integral can be solved whether the elastica knot is closed or not.

Next, for a fixed value for $\gamma = c/(k_{0}D)$, we substitute Eq.~(\ref{eq:lambda_gamma}) into Eq.~(\ref{eq:lambda_mq0}) to obtain
\begin{equation}
    q_{0}(m;\mu) \;=\; \frac{1 + m}{3 + \gamma^{2}} \;\equiv\; \frac{1 + m}{1 + \mu}, \label{eq:q0_m}
\end{equation}
where $\mu = 2 + \gamma^{2} \geq 2$, which then yields
\begin{equation}
    \nu^{2}(m;\mu) \;=\; \frac{(\mu - m)\,(1 - m\,\mu)}{(1 + m)^{2}}. \label{eq:nu2_m}
\end{equation}
Here, the constraint $\nu^{2} \geq 0$ requires that $m \leq 1/\mu < \frac{1}{2}$. 

For the torsionless case $\nu = 0$, which corresponds to planar elastica knot with $m = 1/\mu = q_{0}$, we obtain the periodic {\it cnoidal} solution
\begin{equation}
   \Psi(\xi_{t}) \;=\; k_{0}\;{\rm cn}\left(\sqrt{\mu}\,\xi_{t}|\mu^{-1}\right)\;\exp(i\sqrt{\mu - 2}\,\xi_{t}).
   \label{eq:lemniscate}
\end{equation}
While the Jacobi elliptic function ${\rm cn}(\sqrt{\mu}\,z|\mu^{-1})$ is periodic, with period $\Xi \equiv 4\,{\sf K}(\mu^{-1})/\sqrt{\mu}$, the function $\Psi(\xi_{t})$ acquires a phase shift $\Psi(\xi_{t} + \Xi) = \Psi(\xi_{t})\;\exp[i\Delta\Phi(\mu)]$, where 
\[ 0 \leq \Delta\Phi(\mu) \;=\; 4\sqrt{1 - 2/\mu}\,{\sf K}(\mu^{-1}) \leq 2\pi. \]
Although this phase shift is not necessarily a rational fraction of $2\pi$, the amplitude $|\Psi(s_{t})| = k_{0}\;{\rm cn}\left(\sqrt{\mu}\,\xi_{t}|\mu^{-1}\right)$ is periodic.

Lastly, in the stationary case $(c = k_{0}D\,\gamma = 0)$, obtained when $\mu = 2$, we find $\Psi(\xi) = k_{0}\,
{\rm sn}(\sqrt{2}\,\xi|1/2)$, which corresponds to the so-called leminiscate case \cite{Lawden} (${\sf e}_{1} = -\,{\sf e}_{3}$, ${\sf e}_{2} = 0$, and $\omega_{3} = i\,\omega_{1}$). Appendix \ref{sec:planar} presents the spatial curve corresponding to a planar elastica.

\subsubsection{Weierstrass elliptic solution}

We now wish to show that the introduction of Weierstrass elliptic functions yields a simpler expression for the traveling-wave NLSE solution (\ref{eq:ansatz}), where, using Eq.~(\ref{eq:wp_v}), the phase integral (\ref{eq:Phi_def}) becomes
\begin{equation}
    \theta(\xi_{t}) \;\equiv\; \gamma\,\xi_{t} \;-\; \frac{i}{2}\,\int_{0}^{\xi_{t}} \frac{\wp^{\prime}(v + \omega_{3})\;dw}{\wp(w + \omega_{3}) \;-\; \wp(v + \omega_{3})},
\end{equation}
which has an identical form as Eq.~(\ref{eq:varphi_sol_2}).

Next, we use the identity (\ref{eq:wp_zeta_sigma}) so that we obtain the phase integral
\begin{eqnarray}
    \theta(\xi_{t}) &=& \gamma\,\xi_{t} \;+\; \frac{i}{2}\;\ln\left[\frac{\sigma(\xi_{t}-v)\;\sigma(v + 2\omega_{3})}{\sigma(\xi_{t}+v + 2\omega_{3})\;\sigma(-v)}\right] \;+\; i\;\zeta(v + \omega_{3})\,\xi_{t}.
\end{eqnarray}
Using the quasi-periodicity relation (\ref{eq:sigma_period}) for the odd-parity sigma function, we find
\[ \frac{\sigma(\xi_{t}-v)}{\sigma(\xi_{t}+v + 2\omega_{3})} \;=\; \frac{\sigma(v - \xi_{t})}{\sigma(v + \xi_{t})}\; \exp\left[-\,2\,\eta_{3}\,(\xi_{t} + v + \omega_{3})\right], \]
and
\[ \frac{\sigma(v + 2\omega_{3})}{\sigma(-v)} \;=\; \exp\left[2\,\eta_{3}\,(v + \omega_{3})\right]. \]
Hence, the phase integral is expressed as
\begin{eqnarray}
    \theta(\xi_{t}) &=& \gamma\,\xi_{t} \;+\; \frac{i}{2}\,\ln\left[\frac{\sigma(v - \xi_{t})}{\sigma(v + \xi_{t})}\right] \;+\; i\,\left[ \zeta(v + \omega_{3}) - \zeta(\omega_{3})\right]\,\xi_{t},
\end{eqnarray}
and the traveling-wave NLSE solution (\ref{eq:ansatz_sol}) becomes
\begin{eqnarray}
   \Psi(\xi_{t}) &=& k_{0}\;\sqrt{[\wp(v + \omega_{3}) - \wp(\xi_{t} + \omega_{3})]\;\frac{\sigma(v + \xi_{t})}{\sigma(v - \xi_{t})}} \;\exp\left( i\gamma\,\xi_{t} - i\,[\zeta(v + \omega_{3}) - \zeta(\omega_{3})]\xi_{t}\right).
\end{eqnarray}
This expression can be simplified further by using the identity
\[ \wp(p) - \wp(q) \;\equiv\; \frac{\sigma(p + q)\,\sigma(q - p)}{\sigma^{2}(p)\,\sigma^{2}(q)}, \]
so that we find
\begin{eqnarray} 
\frac{[\wp(v + \omega_{3}) - \wp(\xi_{t} + \omega_{3})]}{\sigma(v - \xi_{t})} &=& -\;\frac{\sigma(v + \xi_{t} + 2\omega_{3})}{\sigma^{2}(v + \omega_{3})\;
\sigma^{2}(\xi_{t} + \omega_{3})} \;=\; \frac{\sigma(v + \xi_{t})\;\exp[2\eta_{3}\,(v + \xi_{t} + \omega_{3})]}{\sigma^{2}(v + \omega_{3})\;\sigma^{2}(\xi_{t} + \omega_{3})}.
 \end{eqnarray}
 We can also use the identity
 \begin{eqnarray*} 
 \wp(v + \omega_{3}) - \wp(\omega_{3}) &=& ({\sf e}_{3} + 1) - {\sf e}_{3} \;=\; 1 \;=\; \frac{\sigma^{2}(v)\;\exp[2\eta_{3}\,(v + \omega_{3})]}{\sigma^{2}(v + \omega_{3})\;\sigma^{2}(\omega_{3})} 
  \end{eqnarray*}
  which implies that
  \[ \sigma(v + \omega_{3}) \;=\; \frac{\sigma(v)}{\sigma(\omega_{3})}\;\exp[\eta_{3}\,(v + \omega_{3})] \]
 The final Weierstrass elliptic expression for the traveling-wave NLSE solution (\ref{eq:ansatz_sol}) can, therefore, be expressed as
 \begin{equation}
    \Psi(\xi_{t}) \;\equiv\; k_{0}\; \left[\frac{\sigma(\xi_{t}+ v)\,\sigma(\omega_{3})}{\sigma(\omega_{3} - \xi_{t})\,\sigma(v)}\right]\;e^{\left[i\gamma - \zeta(v + \omega_{3})\right]\,\xi_{t}},
 \end{equation}
 where we used the identity 
 \[ \sigma(\xi + \omega_{3}) = \sigma(\omega_{3} - \xi)\,\exp(2\eta_{3}\,\xi), \]
 and we easily verify that $\Psi(0) = k_{0}$.

\begin{figure}
\epsfysize=2.1in
\epsfbox{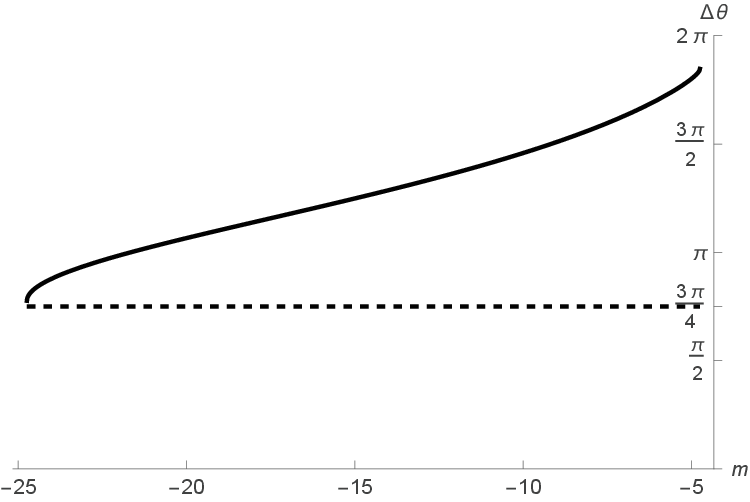}
\caption{Plot of $\Delta\theta(m) = -i \ln[\Psi(\xi_{t} + 2\omega_{1})/\Psi(\xi_{t})]$ as a function of $m$ in the range $m_{1} < m < m_{0}^{-}$. For all values in that range, we find $3\pi/4 < \Delta\theta(m) < 2\pi$.}
\label{fig:Delta_theta}
\end{figure}

Finally, after a single knot period $2\omega_{1}$, we obtain the relation
\begin{eqnarray}
    \Psi(\xi_{t} + 2\omega_{1}) &=& k_{0}\; \left[\frac{\sigma(\xi_{t}+ v + 2\omega_{1})\,\sigma(\omega_{3})}{\sigma(\omega_{3} - \xi_{t} - 2\omega_{1})\,\sigma(v)}\right]\;e^{[i\gamma - \zeta(v + \omega_{3})](\xi_{t}+ 2\omega_{1})} \;=\; \Psi(\xi_{t})\;\exp(i\;\Delta\theta),
    \label{eq:Psi_NLSE_period}
\end{eqnarray}
where the phase shift $\Delta\theta(m) = -i \ln[\Psi(\xi_{t} + 2\omega_{1})/\Psi(\xi_{t})]$ is defined as
\begin{eqnarray}
   \Delta\theta(m) &=& 2\,\gamma\,\omega_{1} + 2\,i\;\left[ \zeta(v + \omega_{3})\;\omega_{1} - \zeta(\omega_{1})\;(v + \omega_{3})
   \right] \nonumber \\
   &=& -\; 2\,\int_{0}^{iv}\left[\omega_{1}\,\wp(\omega_{3} - i\,t) \;+\; \zeta(\omega_{1})\right]\,dt \;+\; 2\,\gamma\,\omega_{1} \;+\; \pi,
\end{eqnarray}
which is shown in Fig.~\ref{fig:Delta_theta}. Here, we easily see that a periodic solution of the NLSE, with the phase shift $\Delta\theta(m)$ equal to a rational fraction of $\pi$, will generically not be on a closed 3D elastica curve, i.e., the azimuthal shift $\Delta\varphi$, given by Eq.~(\ref{eq:Delta_phi}), will generically not be simultaneously the same rational fraction of $\pi$. On the other hand, since only the magnitude $|\Psi(\xi_{t})|$ is physically meaningful, the periodicity of $|\Psi(\xi_{t})|$ is guaranteed by Eq.~(\ref{eq:Psi_NLSE_period}).

\section{\label{sec:sec_8}Conclusions}

In the present work, we investigated how a change of notation for the Jacobi elliptic functions and complete elliptic integrals could open up the elastica parameter $(m,q_{0})$ range for closed 3D elastica curves. The constraint of a closed 3D elastica curve yielded a periodicity condition $\Delta z = 0$ on the vertical position of a point on the curve that introduced the constraint $q_{0} = Q(m)$, given by Eq.~(\ref{eq:q_LS}).

When we explored this extended parameter range, we found that it was possible to obtain a traveling-wave solution for the nonlinear Schr\"{o}dinger equation (\ref{eq:NLSE}), provided the elastica-knot parameters $m \leq m_{0}^{-} < 0$ and $q_{0} = Q(m) < 0$ are in the extended range, which were completely ignored in the classical theory of elastica knots \cite{Langer_Singer_1984,LS_1984}. Further investigation of these traveling-wave NLSE solutions, which are consistent with the Hasimoto transformation (\ref{eq:psi_def}) will be conducted in future work.

\appendix 

\section{\label{sec:Jacobi}Jacobi Elliptic Identities}

The present work uses the conventional notation \cite{AS} ${\rm sn}(z|m)$ for Jacobi elliptic functions, where the argument $z$ may be complex valued while the parameter $m \leq 1$ may be negative. Here, the Jacobi parameter range $0 < m < 1$ is called the classical range, while the range $m < 0$ is called the extended range. In contrast to the mathematical notation \cite{Lawden}, where ${\rm sn}(x,p) \equiv {\rm sn}(x|p^{2})$ in the classical range $0 < m = p^{2} < 1$, the extended range $m < 0$ can be explored continuously without the need for an imaginary parameter $p = \sqrt{m} = i\,\sqrt{|m|}$. 

The purpose of the present Appendix is to introduce the conventional notation \cite{AS} for complete elliptic integrals and Jacobi elliptic functions.

\subsection{Complete elliptic integrals}

\begin{figure}
\epsfysize=2.5in
\epsfbox{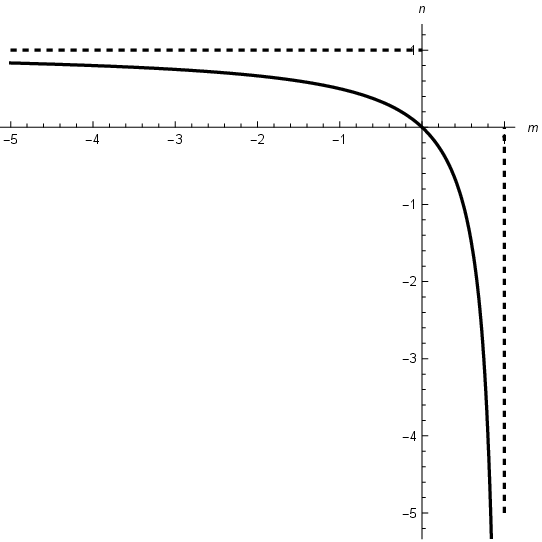}
\caption{Plot of $n(m) \equiv -m/(1 - m)$ versus $m$ in the range $-5 < m < 1$. For each classical range $0 < m < 1$ or $0 < n < 1$, there corresponds a semi-infinite extended range $-\infty < n < 0$ or $-\infty < m < 0$.}
\label{fig:nm_plot}
\end{figure}

First, we begin with the complete elliptic integral of the first kind \cite{AS}
\begin{equation}
    {\sf K}(m) \;=\; \int_{0}^{\pi/2}\;\frac{d\phi}{\sqrt{1 - m\,\sin^{2}\phi}},
    \label{eq:K_m}
\end{equation}
which is real for $0 \leq m < 1$, where ${\sf K}(0) = \pi/2$ and ${\sf K}(m)$ diverges as $m \rightarrow 1$. For $m < 0$, after performing the change of integration variable $\phi = \pi/2 - \theta$, we find the transformation \cite{AS}
\begin{eqnarray}
    {\sf K}(m) &=& \frac{1}{\sqrt{1 + |m|}}\int_{0}^{\pi/2}\;\frac{d\theta}{\sqrt{1 - [|m|/(1 + |m|)]\,\sin^{2}\theta}} \;\equiv\; \sqrt{n'}\;{\sf K}(n),
    \label{eq:K_n}
\end{eqnarray}
where the transformed Jacobi parameters $(n, n' \equiv 1 - n)$ are
\begin{eqnarray}
    n(m) &=&  -m/(1 - m) \;\equiv |m|/(1 + |m|), \label{eq:n_m} \\
    n'(m) &=&  1/(1 - m) \;\equiv 1/(1 + |m|), \label{eq:np_m} 
\end{eqnarray}
which both fall in the classical range $0 \leq n = 1 - n' \leq 1$. Figure \ref{fig:nm_plot} shows a plot of the parameters $n(m)$ versus $m$ in the range $-5 < m < 1$, which demonstrates the complete symmetry between the Jacobi parameters $m$ and $n$, i.e., for each negative parameter $m$ (or $n$), there corresponds a unique classical parameter $n$ (or $m$). For example, the parameter $m = -1$ corresponds to $n = 1/2$ and vice versa.

\begin{figure}
\epsfysize=2in
\epsfbox{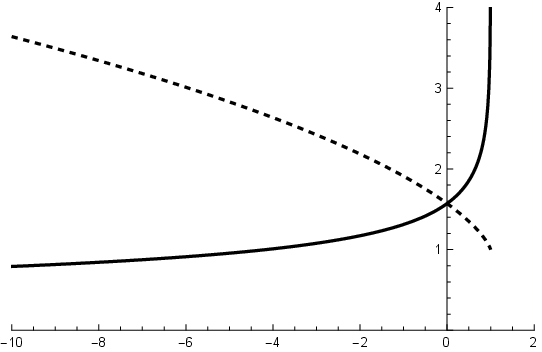}
\caption{Plots of ${\sf K}(m)$ (solid) and ${\sf E}(m)$ (dashed) versus $m$ in the range $-10 < m < 1$. While ${\sf K}(m)$ diverges as $m \rightarrow 1$, we find ${\sf E}(1) = 1$ and ${\sf E}(0) = {\sf K}(0) = \pi/2$.}
\label{fig:EK_m}
\end{figure}

The complete elliptic integral of the second kind, on the other hand, is defined as \cite{AS}
\begin{equation}
    {\sf E}(m) \;=\; \int_{0}^{\pi/2}\;\sqrt{1 - m\,\sin^{2}\phi}\,d\phi,
    \label{eq:E_m}
\end{equation}
which is real for $0 \leq m < 1$, where ${\sf E}(0) = \pi/2$ and ${\sf E}(1) = 1$. For $m < 0$, we find a transformation similar to Eq.~(\ref{eq:K_n}):
\begin{equation}
    {\sf E}(m) \;\equiv\; {\sf E}(n)/\sqrt{n'}.
     \label{eq:E_n}
\end{equation}
Hence, as Fig.~\ref{fig:EK_m} shows, both complete elliptic integrals ${\sf K}(m)$ and ${\sf E}(m)$ are defined continuously on the real axis $-\infty < m < 1$ when calculated directly from Eqs.~(\ref{eq:K_m}) and (\ref{eq:E_m}), respectively. Here, we see that, in the extended range $m < 0$, the complete elliptic integral ${\sf K}(m)$ decreases as $1/\sqrt{1-m}$, while the complete elliptic integral ${\sf E}(m)$ increases as $\sqrt{1-m}$ when $m$ is a large negative number.

We now apply the transformations (\ref{eq:K_n}) and (\ref{eq:E_n}) to explore the behavior of the function (\ref{eq:q_LS}) defined in the classical range $0 < m < 1$:
\begin{equation}
    Q(m) \;=\; 2\,{\sf E}(m)/{\sf K}(m) \;-\; (1 - m),
    \label{eq:Q_m}
\end{equation}
which appears as a result of the periodicity condition on the vertical position on the spatial curve of an elastica knot. For $m < 0$, this function transforms to
\begin{equation}
    Q(m) \;=\; \frac{1}{n'(m)}\left(2\,\frac{{\sf E}(n)}{{\sf K}(n)} \;-\; 1\right),
    \label{eq:Q_n}
\end{equation}
On the one hand, if we strictly use Eq.~(\ref{eq:Q_m}) to find the zeros of $Q(m)$, we find $m = 1$, since ${\sf K}(m) \rightarrow \infty$ as $m \rightarrow 1$, and $m \simeq -4.751$. On the other hand, since the second root is in the extended range $m < 0$, we use Eq.~(\ref{eq:Q_n}) to find the corresponding root at $2\,{\sf E}(n) = {\sf K}(n)$, located in the classical range $n \simeq 0.8261 < 1$, which corresponds exactly to $m = -n/(1 - n) \simeq -4.751$. In fact, the classical parameter $m \simeq 0.8261 < 1$ also corresponds to the root associated with the torsionless case $Q(m) = m < 1$.

Finally, we return to the half-periods $\omega_{1}(m)$ and $\omega_{3}(m)$ of the Weierstrass solution $\wp(z;{\sf g}_{2},{\sf g}_{3})$. While $\omega_{1}(m) = \sqrt{Q(m)}\;{\sf K}(m)$ and $\omega_{3}(m) = i\,\sqrt{Q(m)}\;{\sf K}(1-m)$ when $0 \leq m \leq \frac{1}{2}$ and $Q(m) > 0$, we find $\omega_{1}(m) = \sqrt{|Q(m)|n'(m)}\;{\sf K}[n'(m)]$ and 
$\omega_{3}(m) = i\,\sqrt{|Q(m)|n'(m)}\;{\sf K}[n(m)]$, where $m < m_{0}^{-}$ and $Q(m) < 0$.

\subsection{Jacobi elliptic functions}

Next, we consider the transformations (\ref{eq:transform_J}) of the Jacobi elliptic functions ${\rm sn}(z|m)$, ${\rm cn}(z|m)$ and ${\rm dn}(z|m)$. When $m < 0$ and $z = x$ is real (i.e., $q_{0} > 0$), we use the identities \cite{AS}
\begin{eqnarray}
    {\rm sn}(x|m) &=& \sqrt{n'}\;{\rm sd}(x/\sqrt{n'}|n), \label{eq:sn_x} \\
    {\rm cn}(x|m) &=& {\rm cd}(x/\sqrt{n'}|n), \label{eq:cn_x} \\    
    {\rm dn}(x|m) &=& {\rm nd}(x/\sqrt{n'}|n), \label{eq:dn_x}
\end{eqnarray}
to obtain functions that are periodic, with periods $4\sqrt{n'}\,{\sf K}(n)$ for ${\rm sd}$ and ${\rm cd}$, and $2\sqrt{n'}\,{\sf K}(n)$ for ${\rm nd}$. When $m < 0$ and $z = iy$ is imaginary (i.e., $q_{0} < 0$), on the other hand, we use the identities \cite{AS}
\begin{eqnarray}
    {\rm sn}(iy|m) &=& i\,\sqrt{n'}\;{\rm sd}(y/\sqrt{n'}|n'), \label{eq:sn_y} \\
    {\rm cn}(iy|m) &=& {\rm nd}(y/\sqrt{n'}|n'), \label{eq:cn_y} \\    
    {\rm dn}(iy|m) &=& {\rm cd}(y/\sqrt{n'}|n'), \label{eq:dn_y}
\end{eqnarray}
to obtain functions that are periodic, with periods $4\sqrt{n'}\,{\sf K}(n')$ for 
${\rm sd}$ and ${\rm cd}$, and $2\sqrt{n'}\,{\sf K}(n')$ for ${\rm nd}$. 

Finally, we note that {\sf Mathematica} can seamlessly calculate and plot elliptic integrals and Jacobi elliptic functions over the range $-\infty < m < 1$, i.e., ${\rm sn}(x|m)$ can be plotted continuously over the entire range for $m$, without the need of using the transformation (\ref{eq:sn_x}) when $m < 0$.

\section{\label{sec:equivalent}Curvature Equivalent Elastica Knots}

In this Appendix, we briefly summarize the work of Brizard and Pfefferl\'{e} \cite{AJB_DP}, where an equivalence class of elastica knots is constructed based on either the averaged total curvature or the averaged total torsion, for which an elastica knot with parameters $(q_{0} > 0, m^{-} \leq 0)$ is said to be equivalent to an elastica knot with parameters $(q_{0} > 0,m^{+} \geq 0)$, which are related by the transformation $m^{+} \equiv -\,m^{-}/(1 - m^{-}) < 1$. 

We begin the curvature functional \eqref{eq:F_Lambda}, with the constraint $|{\bf r}^{\prime}| = 1$ now implemented. Inserting the squared-curvature solution
(\ref{eq:elastica_J}), we evaluate the normalized curvature functional
\begin{eqnarray}
\ov{\cal F}(m,q_{0}) = \frac{1}{2k_{0}\,\wh{\kappa}}\int_{0}^{S}\kappa^{2}(s)\;ds & = & \frac{1}{\sqrt{q_{0}\wh{\kappa}^{2}}} \left[ \int_{0}^{2\,{\sf K}(m)}{\rm dn}^{2}(\xi |m)\,d\xi - (1 - q_{0})\;2\,{\sf K}(m) \right] \nonumber \\
   & = & \frac{2}{\sqrt{q_{0}\wh{\kappa}^{2}}} \left[ {\sf E}(m) \;-\frac{}{} (1 - q_{0})\,{\sf K}(m)\right],
\label{eq:F_S}
  \end{eqnarray}
where $q_{0} > 0$ and the normalizing factor
\begin{equation}
q_{0}\,\wh{\kappa}^{2} \;\equiv\; \left\{ \begin{array}{lr}
q_{0} & (m > 0) \\
 & \\
q_{0} - m & (m < 0)
 \end{array} \right. 
 \label{eq:kappa_factor}
 \end{equation}
 guarantees that the curvature functional is normalized with respect to the maximum curvature.

 \begin{figure}
\epsfysize=2in
\epsfbox{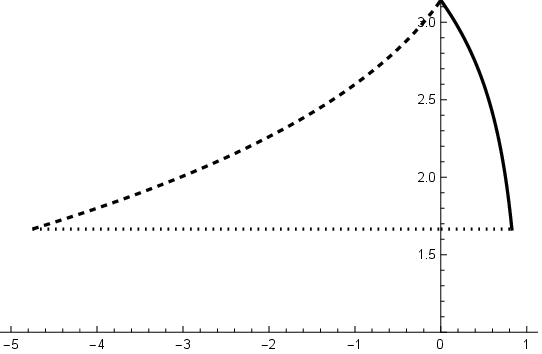}
\caption{Plot of the normalized curvature functional $\wh{\cal F}(m) \equiv \ov{\cal F}(m,Q(m))$ versus $m$ in the range $0 \leq m \leq m_{0}^{+} < 1$ (solid curve) and $m_{0}^{-} \leq m \leq 0$ (dashed curve) for $q_{0} = Q(m)$. At the boundary $m = 0$ ($q_{0} = 1$), we find $\wh{\cal F}(0) = \pi$, while $\wh{\cal F}(m_{0}^{\pm}) > \pi/2$ (dotted horizontal line) at the end points $m = m_{0}^{\pm}$.}
\label{fig:curv_knot}
\end{figure}

Using the constraint $q_{0} = Q(m)$, Eq.~\eqref{eq:F_S} yields 
\begin{equation}
    \wh{\cal F}(m) \;\equiv\; \ov{\cal F}(m,Q(m)),    
\end{equation} 
which is shown in Fig.~\ref{fig:curv_knot} (as a solid curve for $m > 0$ and a dashed curve for $m < 0$). Here, the limiting points $m_{0}^{\pm}$ are defined from the conditions
$Q(m_{0}^{-}) = 0$ at $m_{0}^{-} = -4.751...$ and $Q(m_{0}^{+}) = m_{0}^{+}$ at $m_{0}^{+} = 0.8261...$ At the boundary $m = 0$, we find $\wh{\cal F}(0) = \pi$. Lastly, we note that the normalized curvature functional satisfies the modulus symmetry
\begin{equation}
\wh{\cal F}(n(m)) \;=\; \wh{\cal F}(m),
\end{equation}
with $\wh{\cal F}(m_{0}^{\pm}) = (2 m_{0}^{-} - 1)\,{\sf K}(m_{0}^{-})/\sqrt{m_{0}^{-}} > \pi/2$ (shown as a dotted horizontal line in Fig.~\ref{fig:curv_knot}). For example, we find two equivalent Jacobi elliptic solutions with moduli $m^{+} = 1/2$ and $m^{-} = -\,1$, which have the same numerical values for the normalized total curvature $\wh{\cal F}(-1) = 2.59818... = \wh{\cal F}(1/2)$. We note, however, that their corresponding spatial curves are very different.

\section{\label{sec:planar}Traveling-wave NLSE on an open planar elastica}

The case of zero torsion $\nu = 0 = \Omega$, with $C = 0$ in Eqs.~(\ref{eq:rho_prime})-(\ref{eq:z_prime}), was studied by Hasimoto \cite{Hasimoto_1971}. By inserting $\beta = 0$ into Eqs.~(\ref{eq:t_ab})-(\ref{eq:b_ab}), we obtain $\wh{\sf t} = \sin\alpha\,\wh{\rho} + \cos\alpha\,\wh{\sf z}$, $\wh{\sf n} = -\,\cos\alpha\,\wh{\rho} + \cos\alpha\,\wh{\sf z}$, and $\wh{\sf b} = -\,\wh{\varphi}$, which is a constant vector since $\varphi^{\prime} = 0$. Hence, the torsionless case corresponds to a planar elastica curve on the $(\rho,z)$-plane at constant $\varphi$, which can be chosen to be $\varphi = 0$. 

\begin{figure}
\epsfysize=1.7in
\epsfbox{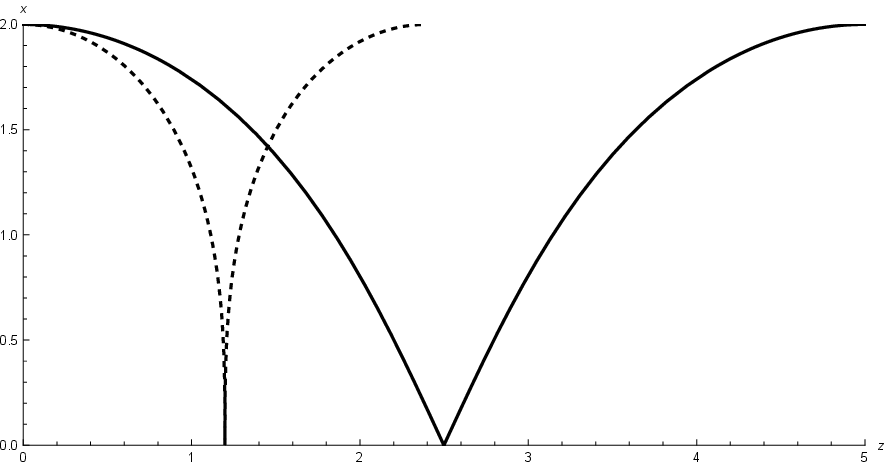}
\caption{Parametric plot of $\ov{\rho}(\xi)$ versus $\ov{z}(\xi)$ for $q_{0} = m$, with $m = 1/4$ (solid) and $m = 1/2$ (dashed), in the range $0 \leq \xi \leq 2\sqrt{m}\,{\sf K}(m)$.}
\label{fig:planar}
\end{figure}

For the torsionless case, we find ${\cal R}^{2}k_{0}^{2} = 2$ and the traveling-wave speed $c(m,q_{0}) = k_{0}D\,\gamma(m,q_{0})$ is determined from Eq.~(\ref{eq:lambda_mq0}):
\begin{equation}
    c(m,q_{0}) \;=\; k_{0}D\;\sqrt{\frac{(1 + m)}{q_{0}} - 3}.
    \label{eq:c_mq0}
\end{equation}
Equation (\ref{eq:nu2_mq0}), on the other hand, implies that $\nu(m,q_{0}) = 0$ either for $q_{0} = 1$ or $q_{0} = m$. For the case $q_{0} = 1$, the normalized vertical and radial solutions (\ref{eq:z_sol}) and (\ref{eq:rho_sol}) are
\begin{eqnarray*}
    \ov{\rho}(\xi) &=& 2\;{\rm dn}(\xi|m), \\
    \ov{z}(\xi) &=& \left[m + 2\,\left({\sf E}(m)/{\sf K}(m) - 1\right)\right]\,\xi + 2\,{\cal Z}(\xi|m),
\end{eqnarray*}
but this case is forbidden as a traveling-wave solution since the wave speed $c(m,1) = k_{0}D\,\sqrt{m - 2}$ is imaginary. For the case $q_{0} = m$, on the other hand, the normalized vertical and radial solutions (\ref{eq:z_sol}) and (\ref{eq:rho_sol}) are
\begin{eqnarray}
    \ov{\rho}(\xi) &=& 2\;|{\rm cn}(\xi/\sqrt{m}|m)|, \label{eq:rho_2} \\
    \ov{z}(\xi) &=& \left[1 + 2\,\left({\sf E}(m)/{\sf K}(m) - 1\right)\right]\,\xi/m \;+\; (2/\sqrt{m})\,{\cal Z}\left(\xi/\sqrt{m}\;|m\right), \label{eq:z_2}
\end{eqnarray}
which is allowed since the wave speed $c(m,m) = k_{0}D\,\sqrt{(1/m) - 2}$ is real in the range $0 < q_{0} = m < 1/2$. The traveling-wave solution for the NLSE for the torsionless case is given in Eq.~(\ref{eq:lemniscate}). This stationary case $\gamma = 0$ corresponds to the upper-most point of region I in Fig.~\ref{fig:gammanu_2}. 

Figure \ref{fig:planar} shows parametric plots of Eq.~(\ref{eq:rho_2}) versus Eq.~(\ref{eq:z_2}) for the cases $q_{0} = m = 1/4$ (solid) and $q_{0} = m = 1/2$ (dashed), which correspond, respectively, to $\gamma = \sqrt{2}$ and $\gamma = 0$. We note that these plots are examples of Jacobi {\it cycloid} curves, in analogy with the standard cycloid curves. At the half-period $\xi = \Xi/2 \equiv \sqrt{m}\,{\sf K}(m)$, for Eqs.~(\ref{eq:rho_2})-(\ref{eq:z_2}), the slope of the curve $d\ov{\rho}/d\ov{z} \equiv \rho^{\prime}/z^{\prime} = \tan\alpha$ is defined by the mid-point angle 
\begin{equation} 
\alpha(m) \;=\; \tan^{-1}\left(\frac{2\sqrt{m\,(1 - m)}}{(1-2m)}\right) \;\leq\; \pi/2. 
\end{equation}
In Fig.~\ref{fig:planar}, we find the mid-point angle $\alpha = \pi/2$ for $m = 1/2$ (dashed curve), and $\alpha = \pi/3$ for $m = 1/4$ (solid curve). Finally, after one period $\Xi$ along the $\xi$-axis, the jump in the vertical position (\ref{eq:z_2}) is equal to
\begin{equation}
    \Delta\ov{z}(m) \equiv \ov{z}(\xi + \Xi) - \ov{z}(\xi) = \frac{2}{\sqrt{m}}\;\left[ 2\,{\sf E}(m) - {\sf K}(m)\right],
\end{equation}
which diverges as $m \rightarrow 0$, while it reaches the limiting value $\sqrt{2}\pi/{\sf K}(1/2) \simeq 2.396$ at $m = 1/2$.

\end{document}